\documentclass[12pt]{iopart}

\usepackage{graphicx}% Include figure files
\usepackage{dcolumn}% Align table columns on decimal point
\usepackage{bm}% bold math
\usepackage{bbm}
\usepackage{epsfig}
\usepackage{mathrsfs}
\usepackage{stmaryrd}
\usepackage{color}
\usepackage{dsfont}
\usepackage{iopams}
\usepackage{psfrag}

\newcommand{\openone}{\bf{1}}
\newcommand{\bra}[1]{\langle\,{#1}\, |}
\newcommand{\ket}[1]{|\,{#1}\,\rangle}
\newcommand{\braket}[2]{\mbox{$\langle\,{#1}\, | \,{#2}\,\rangle$}}

\newcommand{\Real}{\mbox{Re}}

\newcommand{\vek}[1]{\boldsymbol{#1}}

\newcommand{\nextrow}{\vspace{2mm}\\}
\newcommand{\HamTot}{H}
\newcommand{\HamEl}{H^{\rm el}}
\newcommand{\Vdd}{V}

\newcommand{\Ort}{R}

\newcommand{\Pos}{\Ort}
\newcommand{\PosAbs}{\Ort}

\newcommand{\atomA}{2}
\newcommand{\atomB}{3}
\newcommand{\ssection}[1]{{\noi  \it #1:}}
\newcommand{\pdiff}[2]{\frac{\partial #1}{\partial #2}}

\newcommand{\sub}[2]{{#1}_{ \mbox{\scriptsize #2}}}

\newcommand{\bv}[1]{\mathbf{ #1 }}

\def\noi{\noindent}
\def\beq{\begin{equation}}
\def\eeq{\end{equation}}

\def\CR{\nonumber\\[0.15cm]}

\newcommand{\rref}[1]{Ref.~\cite{#1}}
\newcommand{\frefp}[2]{Fig.~\ref{#1}~(#2)}
\newcommand{\aref}[1]{\ref{#1}}
\newcommand{\bref}[1]{(\ref{#1})}
\usepackage{ulem}  %unter- (\uline{important}) und durchstreichen (\sout{wrong})
\begin{document}

%%%%%%%%%%%%%%%%%%%%%%%%%%%%%%%%%%%%%%%%%%%%%%%%%%%%%%%%%%%%%%%%%

\title{Adiabatic entanglement transport in Rydberg aggregates}
\author{S.~M\"obius,  S.~W\"uster, C.~Ates, A.~Eisfeld and J.~M.~Rost}
\address{Max Planck Institute for the Physics of Complex Systems, N\"othnitzer Strasse 38, 01187 Dresden, Germany}
\ead{sew654@pks.mpg.de}

%%%%%%%%%%%%%%%%%%%%%%%%%%%%%%%%%%%%%%%%%%%%%%%%%%%%%%%%%%%%%%%%%

%\date{\today}% It is always \today, today,
            %  but any date may be explicitly specified

\begin{abstract}
We consider the interplay between excitonic and atomic motion in a regular, flexible chain of Rydberg atoms, extending our recent results on entanglement transport in Rydberg chains [W\"uster {\it et al.}, Phys.~Rev.~Lett {\bf 105} 053004 (2010)].
In such a Rydberg chain, similar to molecular aggregates, an electronic excitation is delocalised due to long range dipole-dipole interactions among the atoms. The transport of an exciton that is initially trapped by a chain dislocation is strongly coupled to nuclear dynamics, forming a localised pulse of combined excitation and displacement. 
This pulse transfers entanglement between dislocated atoms adiabatically along the chain. Details about the interaction and the preparation of the initial state are discussed.
We also present evidence that the quantum dynamics of this complex many-body problem can be accurately described by selected quantum-classical methods, which greatly simplify investigations of excitation transport in flexible chains.

\end{abstract}

%\maketitle
%%%%%%%%%%%%%%%%%%%%%%%%%%%%%%%%%%%%%%%%%%%%%%%%%%%%%%%%%%%%%%%%%
%\section{Introduction}
\section[Introduction]{Introduction\label{introduction}}

Rydberg atoms have recently received increasing attention in cold atomic
physics, to a large part due to their strong long-range interactions, with
diverse consequences from dipole-blockade
\cite{lukin:quantuminfo,urban:twoatomblock,gaetan:twoatomblock} to long range
molecules \cite{Greene:LongRangeMols,liu:ultra_long_range_2009,pfau:rydberg_trimers}. Among the
interactions in cold Rydberg gases, 
resonant dipole-dipole interactions \cite{anderson:resonant_dipole,li_gallagher:dipdipexcit,noordam:interactions}
and their non-resonant variant (van-der-Waals interactions) \cite{amthor:vanderwaals,schempp:poptrap} are particularly prominent. These interactions enable Rydberg ensembles to simulate the quantum dynamics of other long-range interacting systems, from condensed matter physics \cite{lesanovsky:manybodyspin,nils:supersolids} to molecular aggregates \cite{noordam:interactions,MueBlAm07_090601_,cenap:motion}. We focus on the latter possibility, and explore basic consequences of joint dynamics of atomic motion and excitonic transport.

Within an essential state picture, where only two Rydberg states per
 atom are taken into account, the transfer of excitation can be adequately
 described by using the exciton theory of Frenkel \cite{Fr30_198_,Fr31_17_}.
 Following the pioneering paper by Franck and Teller \cite{FrTe38_861_}, this theory has found wide application in describing
 excitation transfer, e.g.\ in molecular crystals \cite{Da62__},
 photosynthesis \cite{AmVaGr00__} or organic dye aggregates \cite{Ko96__}.
In all these systems the coupling between the exciton and nuclear degrees of
freedom strongly influences the excitation transfer
\cite{FrTe38_861_,Bi67_1484_,HaRe71_253_,RoEiWo09_058301_,ReChAs09_184102_,BlSi78_3589_,BaSz87_339_,Ho66_208_}. Similar
effects will be reported in the present study. 

The strong interactions between the monomers of molecular aggregates lead to coherently delocalised entangled states \cite{EiBr02_61_,CaChDa09_105106_,sarovar_fleming:entanglement} which are e.g.~responsible for the J-band of organic dye aggregates. Recent experiments indicate robust excitonic coherence even in biological systems, such as photosynthetic complexes \cite{engel_fleming:coherence_nature, lee_fleming:coherence_science, Collini_scholes:coherence_science}.

In all these excitonic systems the \emph{resonant} nature of the
 interaction plays a crucial role. Besides the transfer of excitation, this
 interaction also creates a potential, which for an atom pair depends like
 $1/R^3$ on their distance $R$.
For Rydberg atoms it has been recognized that this potential can lead to large forces on the individual
atoms \cite{LiTaGa05_173001_,cenap:motion} and thus cause their motion.
In contrast to the atomic motion induced by van-der-Waals  interaction, which is due to
strongly off-resonant coupling, the character of the motion (repulsive, attractive or even mixed) in the resonant case
depends strongly on the excitonic eigenstates \cite{cenap:motion}. These in turn depend on the atomic positions, which is why excitation transport and motion become interlinked.
In this respect our setup \cite{cenap:motion,wuester:cradle} strongly differs from that in Ref.~\cite{asadian:motion}, where the effect of \emph{externally enforced} atomic motion on exciton transport is studied. 

In this article, we extend our previous studies of excitons and their dynamics in Rydberg chains \cite{cenap:motion,wuester:cradle}. To study exciton dynamics with Rydberg ensembles, one requires strong selectivity of the accessible electronic states of each atom and control over the initial exciton state. We consider both requirements, and furthermore provide additional details on the Newton's cradle type entanglement transport scenario reported in \cite{wuester:cradle}. Specifically we vary atomic masses and interaction potentials. For the entanglement transport scenario, we show that two mixed quantum-classical methods are well suited to describe this  complex many-body problem: Tullys surface hopping method, and the Ehrenfest method. For short chains, we validate these quantum-classical propagation methods by comparison with a full quantum mechanical calculation, finding perfect agreement. 

The paper is organized as follows: In \sref{aggregates} we describe which conditions we imply in order to label a Rydberg chain as Rydberg \emph{aggregate}. After a brief comparison with molecular aggregates (\sref{atom_vs_mol}), we describe our geometric setup and Hamiltonian (\sref{general_setup}), illustrate how a simple treatment of  transition dipole-dipole interactions can emerge (\sref{Vdipdip}), argue the validity of our essential states model (\sref{essential}), lay the basis for a description of the Rydberg chain's excitations in terms of excitons and their localisation (\sref{localisation}) and show how the initial states for our later applications could be obtained (\sref{initialstate}). The final part of \sref{aggregates} (\sref{methods}) details the quantum and quantum-classical formalisms used to simultaneously model the dynamics of atomic motion and excitons.
After these preparations, we proceed in \sref{cradle} to a detailed presentation of the entanglement transport scenario first reported in \rref{wuester:cradle} and survey the parameter space for this scenario in \sref{survey}. Some appendices supply further details.

%%%%%%%%%%%%%%%%%%%%%%%%%%%%%%%%%%%
%%%%%%%%%%%%%%%%%%%%%%%%%%%%%%%%%%%
\section{Rydberg aggregates} 
%%%%%%%%%%%%%%%%%%%%%%%%%%%%%%%%%%%
%%%%%%%%%%%%%%%%%%%%%%%%%%%%%%%%%%%
\label{aggregates}

%%%%%%%%%%%%%%%%%%%%%%%%%%%%%%%%%%%
\subsection{Brief comparison with molecular aggregates}
%%%%%%%%%%%%%%%%%%%%%%%%%%%%%%%%%%%
\label{atom_vs_mol}
Since molecular aggregates have been extensively studied over the last 70
years it is appropriate to briefly juxtapose the Rydberg aggregates to
these "conventional" molecular aggregates.

Molecular aggregates appear in various contexts, ranging from organic
crystals \cite{Da62__,ScWo06__} over self-assembled cylindrical dye
aggregates \cite{EiKnKi09_658_} to complex biological light harvesting
systems \cite{AmVaGr00__}.
These systems range from only two monomers up to thousands of monomers, which can aggregate into various geometrical arrangements.
The (resonant) transition-dipole-dipole interaction between the monomers leads to entangled states, often
accompanied by a drastic
change in the absorption spectrum compared to that of the single monomer \cite{Ko96__,EiBr02_61_}. 
Besides some fundamental interest (e.g.\ in photosynthesis) 
the extraordinary  properties of these aggregates have led to various
applications, ranging from sensitisers in photography \cite{Ta96__,TaSuKa06_16169_}, to the measurement of
membrane potentials \cite{ReSmCh91_4480_,DeCoRo01_653_}, and cancer therapy \cite{RyAmPe02_801_}.
Also in the development of efficient, low-cost  artificial light harvesting
units (like organic solar cells) dye aggregates might play an important role
\cite{Dae02_81_,KiDa06_20363_}.

In molecular aggregates the monomers are held at their positions and orientations e.g.\ by a protein environment or by van-der-Waals interactions, with distances of the order of a few \AA ngstr\"om. In the Rydberg aggregates investigated here, the distances are of the
order of a few micrometers and, most importantly, the Rydberg atoms are free
to move.
The main difference, however,  between molecular aggregates and
Rydberg aggregates is the internal structure and the environment.
While the Rydberg atoms are at ultra-cold temperatures and interact only
weakly with the environment, the electronic excitation in the molecular
case does strongly couple to the environment (often at
ambient temperature) and a plethora of internal vibrational modes
\cite{RoEiDv11_054907_}.
This typically necessitates various approximations and assumptions in the theoretical
description of molecular aggregates , since often details about the environment or even the precise arrangement of the monomers are unknown.
Furthermore, due to the small distances, the direct experimental observation of coherent energy transfer in molecular aggregates is challenging.
Hence, related investigations are typically of spectroscopic nature \cite{EnCaRe07_782_}, and infer the exciton dynamics only indirectly.

In contrast the beauty of Rydberg aggregates is, that individual excitation
and manipulation of the atoms can be done more easily.
Also, since environment and vibrations do not play a role, it is
possible to develop a detailed theoretical description where common
approximations can be checked.

%%%%%%%%%%%%%%%%%%%%%%%%%%%%%%%%%%%
\subsection{General setup}
%%%%%%%%%%%%%%%%%%%%%%%%%%%%%%%%%%%
\label{general_setup}

%In the following we outline how ensembles of Rydberg atoms can be arranged as an aggregate. 
We study a chain of $N$ identical atoms with mass $M$ and denote the position of the $n$th atom by $\bv{R}_{n}$. These positions are grouped into a $3N$ dimensional vector $\bv{R}=(\bv{\Pos}_1,\dots,\bv{\Pos}_N)^T$\footnote{
We present our introductory theory as far as possible in three dimensions (3D). For all our results we only consider one-dimensional motion (1D), assuming transverse motion is frozen out by the confinement of the atoms.
}. 
In the following we will refer to these coordinates as nuclear coordinates. Each atom should be initially well localised, for example in the ground state of an optical lattice or a micro-lens array \cite{birkl:fortagh:microlensarrays}. We can then ensure that the distance $\PosAbs_{nm}\equiv | \bv{R}_{n}-\bv{R}_{m}|$ between the atoms is large enough to neglect the overlap between the electronic wave functions of atoms $n$ and $m$. 

\begin{figure}[bt]
\centerline{\psfig{file=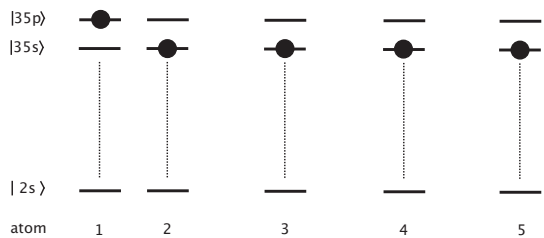,width= 0.45\columnwidth}}
\vspace*{8pt}
\caption{\label{fig:sketch:elec}Visualization of the electronic state $\ket{\pi_1}$.}
\end{figure}
Consider a situation where all but one of the $N$ atoms are in a Rydberg state $|\nu s\rangle$, with principal quantum number $\nu$ and angular momentum $l=0$.
The remaining atom is in an angular momentum $l=1$ state $|\nu p\rangle$, which we will call the ``excited'' state. We now define the single-excitation Hilbert space, whose electronic part is spanned by
\begin{equation}
\label{def:pi_n}
\ket{\pi_n}\equiv \ket{s\cdots p\cdots s},
\end{equation}
a state in which atom $n$ is in the $p$ state and all others are in the $s$ state.
In \fref{fig:sketch:elec} the state $\ket{\pi_1}$ and the initial spatial arrangement is sketched for the case $N=5$. 
Note that the role of s and p is more or less interchangeable. 

For a clear cut picture of exciton transport, the states \bref{def:pi_n} should form the essential part of the electronic basis for the whole aggregate. This requires that transitions to other states, such as $\ket{s\cdots p\cdots s \cdots p\cdots s}$ or $\ket{s\cdots d \cdots s}$ are negligible, because they are energetically far detuned with respect to the relevant couplings. Then, the only relevant interactions occur within the space spanned by \eref{def:pi_n} and conserve the number of excitations. In \sref{essential} we show by an example that these requirements can be fulfilled.

In terms of the basis \eref{def:pi_n}, our total Hamiltonian describing atomic motion and interactions within the essential states manifold is given in atomic units by
\begin{equation}
\label{Ham_Tot}
\HamTot(\bv{\Pos})=-\sum_{n=1}^N\frac{\nabla^2_{\bv \Pos_n}}{2 M}  + H^{\rm el}(\bv{\Pos}).
\end{equation}
Here, the electronic Hamiltonian, which depends on the nuclear coordinates, is
\begin{equation}
\label{H_el}
\HamEl(\bv{\Pos})=\sum_{nm}\Vdd_{nm}(\PosAbs_{nm})\ket{\pi_n}\bra{\pi_m},
\end{equation}
where 
\begin{equation}
\Vdd_{nm}(\Pos_{nm})=(-1)^{\eta}\frac{\mu^2}{\PosAbs_{nm}^3}
\end{equation}
is the dipole-dipole coupling between atoms $n$ and $m$ and $\PosAbs_{nm}$ their separation. We parameterised the strength of the coupling by its magnitude $\mu^{2}$ and sign $\eta \in\{0,1\}$. Due to this resonant dipole-dipole interaction the ``excitation'' can be transferred from atom $n$ to $m$. 
We outline in \sref{Vdipdip} why we can avoid a more complicated, angular dependent \cite{noordam:interactions} expression.

For most specific examples throughout this article, we consider the atomic species $^{7}$Li Lithium. Among the work horses of cold atom physics, this atom is one of the lightest and hence most suited to display phenomena of dipole-dipole interaction induced motion, within the time-scales available. Its atomic mass is roughly $M=11000\ \rm{ au}$ and transition dipole moment have the strength $\mu=1000\ \rm{ au}$ between $s$ and $p$ states with a principal quantum number $\nu\approx 30 \dots 40$. 

In \sref{survey} we survey the response of dynamics dictated by \eref{Ham_Tot} to changes of Hamiltonian parameters. To this end we also generalise the type of interaction, considering  $\Vdd_{nm}(\Pos_{nm})=(-1)^{\eta}\mu^2/\PosAbs_{nm}^\alpha$, where $\alpha$ can for example vary from $\alpha=1\dots6$, with character of the interaction potentials ranging from Coulombic to van-der-Waals. Keep in mind though, that unlike conventional Coulomb or van-der-Waals interactions, those considered here would still have a resonant transition character.

%%%%%%%%%%%%%%%%%%%%%%%%%%%%%%%%%%%
\subsection{Dipole-dipole interactions}
%%%%%%%%%%%%%%%%%%%%%%%%%%%%%%%%%%%
\label{Vdipdip}

In this section we outline how the simple form $\Vdd_{nm}(\Pos_{nm})=(-1)^{\eta}\mu^2/\PosAbs_{nm}^3$ can be obtained for dipole-dipole interactions.
For this purpose, we consider a binary atom system with separation $\bv{R}_{nm}=\bf{R}_m-\bv{R}_n$ and define $R=|\bv{R}_{nm}|$ and $\hat{\bv{R}}=\bv{R}_{nm}/R$. We assume one of the atoms is in a $|\nu s\rangle$ state and the other in a $|\nu p\rangle$ state, where $\nu$ is the (large) principle quantum number, subsequently suppressed. As long as one ignores directional effects, the essential states Hilbert space for such two atoms is spanned by $|sp\rangle$, $|ps\rangle$. Considering angular dependent \emph{transition} dipole-dipole interactions amounts to taking into account also the magnetic quantum number. 
 We then have six essential states: $|\{p,1\}s\rangle$, $|\{p,0\}s\rangle$, $|\{p,-1\} s\rangle$, $|s\{p,1\}\rangle$, $|s\{p,0\} \rangle$, $|s\{p,-1\}\rangle$, using an obvious notation that writes the magnetic quantum number $m$ of the atom with $l=1$ within the curly brackets.

The non-vanishing dipole-dipole transition amplitudes between those states are \cite{noordam:interactions}
\begin{eqnarray}
V_{1m,00;00,1m'}=&-\sqrt{\frac{8 \pi}{3}}\frac{(d_{\nu 1,\nu 0})^2}{R^3}(-1)^{m'}
\CR
&\times\left( 
\begin{array}{ccc}
1 & 1& 2
\\
m & -m' & m'-m
\end{array}
\right)
Y_{2,m'-m}(\hat{\bv{R}}),
\label{Vhop}
\end{eqnarray}
where $Y_{l,m}$ are spherical harmonics and $(\cdots)$ denotes the Wigner $3j$ coefficient. The matrix-element $V_{l_{1}m_{1},l_{2}m_{2};l'_{1}m'_{1},l'_{2}m'_{2}}$ describes a transition between the two-atom states indicated with primed and non-primed subscripts. $d_{\nu 1, \nu 0}$ is the radial overlap-matrix element between the $l=0$ and $l=1$ states. We refer to \rref{noordam:interactions} for further details.
In matrix-form, using the above basis ordering, one obtains
\begin{eqnarray}
\label{V_mag}
V=\left[  
\begin{array}{cc}
0 & V_{ps} \\
V_{ps}^{\dagger} & 0 \\
\end{array} 
\right]
\end{eqnarray}
with sub-matrices
\begin{eqnarray}
\label{Vps}
&V_{ps}=
\frac{\tilde{\mu}^2}{R^3}\left[  
\begin{array}{ccc}
\frac{3 \cos^2 \theta -1}{6} & \frac{e^{-i \phi}}{\sqrt{2}} \cos\theta \sin\theta & \frac{e^{-2i\phi} \sin^2\theta}{2}
\nextrow
\frac{e^{i \phi}}{\sqrt{2}} \cos\theta \sin\theta  & \frac{1 - 3 \cos^2 \theta}{3}   &   -\frac{e^{-i \phi}}{\sqrt{2}} \cos\theta \sin\theta
\nextrow
\frac{e^{2i\phi} \sin^2\theta}{2} & -\frac{e^{i \phi}}{\sqrt{2}} \cos\theta \sin\theta & \frac{3 \cos^2 \theta -1}{6}
\end{array} 
\right]
\end{eqnarray}
In this matrix, the element $(V_{ps})_{ij}$ contains the amplitude of transitions from a state $|\{p,m_j\}s\rangle$ to $|s\{p,m_i\})\rangle$, where $m_i,m_j \in \{1,0,-1\}$.
We used the short-hand $\tilde{\mu}^2=(d_{n_{a}1,n_{b}0})^2$.
The angles $\theta$ and $\phi$ describe $\hat{\bv{R}}$ in a spherical polar co-ordinate system whose $z$-axis ($\hat{z}$) is the quantization axis with respect to which the magnetic quantum numbers $m$ is defined. A useful choice of $\hat{z}$ will be given by the polarisation direction of the light-field used for the initial-state creation, see \sref{initialstate}.

We will consider two specific simple cases, assuming a linear Rydberg chain.

\ssection{case (i)} Choose $\hat{z}$ along the direction of the chain. Then for all distance vectors $\bv{R}_{nm}$ we have $\theta=0$ and 
\begin{eqnarray}
\label{Vps1}
&V_{ps}=\frac{\tilde{\mu}^2}{3R^3}
\left[  
\begin{array}{ccc}
1 & 0 & 0\\
0 & -2 & 0\\
0 & 0 & 1 \\
\end{array} 
\right].
\end{eqnarray}
Thus the magnetic quantum number of the excitation is conserved. Depending on the selected magnetic quantum number $m$, we can realise different signs $\eta$ and magnitudes of the interaction. 

\ssection{case (ii)}  Choose $\hat{z}$ perpendicular to the direction of the chain, which we assume to be in the $\hat{x}$ direction. We then have (setting $\theta =\pi/2$ and $\phi=0$)
\begin{eqnarray}
\label{Vps2}
&V_{ps}=\frac{\tilde{\mu}^2}{6R^3}
\left[  
\begin{array}{ccc}
-1 & 0 & 3\\
0 & 2 & 0\\
3 & 0 & -1 \\
\end{array} 
\right].
\end{eqnarray}
It can be seen that the $m=0$ state decouples and yields a dipole-dipole interaction transport without angular-dependence.
For all choices of quantisation axis and magnetic quantum state, we finally define the parameter $\mu^2$ used in \sref{general_setup} as the modulus of the factor multiplying $R^{-3}$. How a specific magnetic quantum-number for the excitation can be realised is described in \sref{initialstate}.

%%%%%%%%%%%%%%%%%%%%%%%%%%%%%%%%%%%
\subsection{Validity of essential states model}
%%%%%%%%%%%%%%%%%%%%%%%%%%%%%%%%%%%
\label{essential}

Dipole-dipole transitions in principle do not only take place within the $\ket{sp}$, $\ket{ps}$ manifold, but also to other states not included in our essential states picture \cite{noordam:interactions}. Nonetheless, parameters where these other states can be ignored can easily be found, as we demonstrate now. 

We exploit that transition probabilities from the two original states to all other di-atomic states are negligibly small due to their energy mismatch or selection rules. In other words, all two-atom states to which a direct dipole-dipole transition is possible are much farther detuned than the strength of the transition matrix element. We will illustrate this argument in the following for $^7$Li.

Consider the two-atom states $\ket{35,s}\varotimes \ket{35,p}$, $\ket{35,p}\varotimes \ket{35,s}$. The two-atom states energetically nearest and connected via a single dipole-dipole transition are $\ket{36,s}\varotimes \ket{34,p}$, $\ket{34,p}\varotimes \ket{36,s}$, detuned from our essential states by $\Delta=8.78$ GHz and connected with a coupling strength of about $V=65$ MHz\footnote{
We obtain the lithium state energies using quantum-defect theory as described in \cite{stevens:Lidefects} and transition strength as outlined in \cite{noordam:interactions} with numerical Numerov calculation of radial overlaps.
}. 
A simple analytical four state model for detuned Rabi-oscillations then predicts population transfer out of the essential-state system of the order of $(V/\Delta)^2 = 5\times10^{-5}$ for time scales considered in this paper.

As more rigorous justification of the essential state mode, also accounting for successive, cascaded transitions, 
we propagated the state $\ket{35,s}\varotimes \ket{35,p} + \ket{35,p}\varotimes \ket{35,s}$ (an exciton eigenstate, see \sref{localisation}) with a Hamiltonian
that contains all $\nu$,$l$,$m$ states from $\nu=34$ to $\nu=36$, setting all dipole-allowed
transition matrix elements $V$ to $V=\mu^2/d^3$ for $\mu=1000$ au and $d=2 {\mu}m$. This value for $\mu$ overestimates almost
all transition dipoles and the value for $d$ is the smallest separation occurring in the atomic dynamics of \sref{cradle}.
Within this "worst-case" scenario, total transitions out of our target essential states manifold are of the order of $1\times10^{-4}$ within $20 {\mu}s$, which is longer than the simulated time-span in \sref{cradle}.

Finally kinetic and potential energies of the dynamics presented in our manuscript amount to only about
5\% of the energetic separation between $\ket{35p}$, $\ket{35d}$. 

These estimates, while exemplary, show that there is no general problem with finding physically realistic scenarios with Rydberg atoms
that can be described well with our model. 

%%%%%%%%%%%%%%%%%%%%%%%%%%%%%%%%%%%
\subsection{Excitons, exciton localisation and full aggregate initial state}
%%%%%%%%%%%%%%%%%%%%%%%%%%%%%%%%%%%
\label{localisation}

To gain some insight into the structure of the dynamics induced by the Hamiltonian \bref{Ham_Tot}, consider eigenstates of the electronic Hamiltonian
\begin{eqnarray}
\label{exciton_state}
H^{\rm el}(\bv{R} )\ket{\varphi_{k}(\bv{R})}&=U_{k}(\bv{R} )\ket{\varphi_{k}(\bv{R})}.
\label{adiab_rep}
\end{eqnarray}
For each $\bv{R}$ there are $N$ eigenstates labeled by the index $k$. 
Each of these eigenstates can be expanded in terms of the previously introduced basis $\ket{\pi_n}$ as
\begin{equation}
\ket{\varphi_{k}(\bv{R})} = \sum_{m} c_{km}(\bv{R}) \ket{\pi_m}.
\label{exciton_expansion}
\end{equation}
These eigenstates are termed Frenkel ``excitons'' \cite{Fr30_198_,Fr31_17_} and form an adiabatic (Born-Oppenheimer) basis in the language of molecular physics. The corresponding eigenenergies $U_{k}(\bv{R} )$, which also depend parametrically on the nuclear coordinates   $\bv{R}$, define the adiabatic potential surfaces. As evident from \eref{exciton_expansion}, an exciton is a coherent superposition of different localised excitation states. 

\begin{figure}[bt]
\psfrag{a/x0}{\small $a/x_0$}
\psfrag{ppi1}{\scriptsize $\ket{\pi_1}$}
\psfrag{ppi2}{\scriptsize$\ket{\pi_2}$}
\psfrag{ppi3}{\scriptsize$\ket{\pi_3}$}
\psfrag{ppi4}{\scriptsize$\ket{\pi_4}$}
\psfrag{ppi5}{\scriptsize$\ket{\pi_5}$}
\psfrag{ppi6}{\scriptsize$\ket{\pi_6}$}
\centerline{\psfig{file=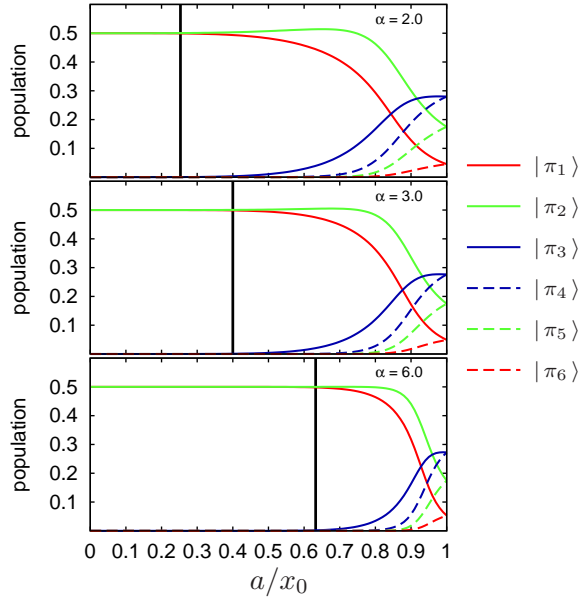, width= 0.5\columnwidth}}
\caption{\label{exciton_localisation}Initial population of the (out of phase) repulsive excitonic state for different $\alpha$ as a function of the dislocation ratio $a/x_0$. The vertical black line 
indicates our choices for $a/x_0$ employed in \sref{survey}. Other lines indicate the population in the different excitonic states for a chain containing 6 Rydberg atoms. }
\end{figure}
\begin{figure}[bt]
%\psfrag{atom}{\small atom nr}
\psfrag{a}{\small $a$}
\psfrag{x}{\small $x_0$}
\psfrag{0}{\ }
\centerline{\psfig{file=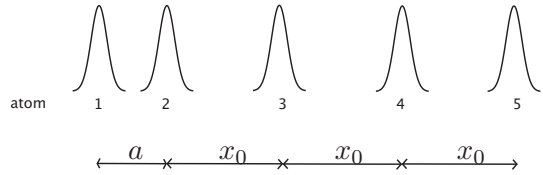,width= 0.45\columnwidth}}
\vspace*{8pt}
\caption{\label{fig:sketchnuc} Sketch of the initial total density distribution of Rydberg atoms for the case of $N=5$ atoms.}
\end{figure}

Consider now a regular chain of Rydberg atoms with spacing $x_{0}$, which is perturbed by a dislocation of two atoms in close mutual proximity (distance $a$), see \fref{fig:sketchnuc}. The interaction between these atoms is much larger than interactions involving the remaining atoms. As a consequence two of the exciton states localise on the dislocation atoms. For $a\ll x_0$ the state whose Born-Opp.\ surface has repulsive character \cite{cenap:motion} can be approximately written as $ \ket{\varphi_{\rm rep}}\approx(\ket{\pi_1}+ (-1)^{\eta}\ket{\pi_2})/\sqrt{2}$.
Such repulsive dimer states are observed e.g.~in \cite{li_gallagher:dipdipexcit}.
In \fref{exciton_localisation} the excitonic population on the various atoms as a function of $a/x_0$ is shown for the case $N=6$ and different interaction exponents $\alpha$. 
Our survey of dynamics presented later requires sufficiently good coherent exciton localisation on the dislocation atoms, which lead to our choices of $a/x_{0}$ indicated by the vertical black lines in the figure.

We now are in position to discuss the whole initial state of our Rydberg aggregate, describing the electronic state and the position of the atoms. The initial \emph{spatial} wave function of each atom is assumed Gaussian with standard deviation $\sigma_{0}$. This resembles an experimental situation where the ground-state atoms are trapped in harmonic potentials prior to their excitation to the Rydberg level, as discussed in the next section. For a sketch of this arrangement, see \fref{fig:sketchnuc}.

We take the complete initial wave function (i.e.\ containing nuclear and
excitonic degrees of freedom) as
\begin{eqnarray}
\label{psi_tot_ini}
\ket{\Psi(t=0)}&=\ket{\varphi_{\rm rep}(\bv{R})}  \prod_{n=1}^N \phi_{\rm G}(\bv \Pos_n),
\\
\phi_{\rm G}(\bv \Pos_n)&={\cal N}\exp{(-|\bv \Pos_n-\bv \Pos_{0n}|^2/2 \sigma_0^2 )},
\end{eqnarray}
where $\Pos_{0n}$ is the center of mass of the $n$-th Gaussian and ${\cal N}$ a normalization factor. As above, the index ``rep'' denotes the unique exciton state with globally repulsive behaviour \cite{cenap:motion}. 
In \eref{psi_tot_ini}, the exciton state varies with the nuclear co-ordinates $\bv{R}$ within the initial Gaussian distribution, but as long as $\sigma_{0}\ll a$ this effect will be small.
We found that one obtains almost identical dynamics to that presented in the following, if $\varphi_{\rm rep}(\bv{R})$ is replaced by $\varphi_{\rm rep}(\bv{R_0})$ in \eref{psi_tot_ini}.

%%%%%%%%%%%%%%%%%%%%%%%%%%%%%%%%%%%
\subsection{State preparation}
%%%%%%%%%%%%%%%%%%%%%%%%%%%%%%%%%%%

\label{initialstate}

In this section we will briefly describe how states discussed in the previous section could be prepared experimentally.

At the very beginning, ground state atoms are confined e.g.~in optical traps created by microlens arrays that provide the desired spacial arrangement, in our case a linear chain with distances $a$ and $x_0$.
Ideally there would be exactly one atom per trap-site.
These ground state atoms are then transferred into a certain Rydberg state (say 35s) via laser excitation\footnote{If there is more than one atom in the ground state trap one can ensure, due to dipole-blockade, that there is actually only one Rydberg excitation per site.}. 
We now have a state, which we denote by $\ket{s\cdots s}$, where all atoms are in the same Rydberg state $\ket{35s}$. 
Due to the ultra-cold temperatures the Rydberg atoms can be regarded as frozen (the distances between the sites have to be chosen such that acceleration and blockade effects due to van-der-Waals interaction \cite{AmReWe07_023004_} are negligible).   
Then by applying a short microwave pulse, which will be specified below, the eigenstate of the chain $\ket{\psi_{\rm ini}}\approx (\ket{\pi_1}+ \ket{\pi_2})/\sqrt{2}$ can be excited. To reach $\ket{\psi_{\rm ini}}\approx (\ket{\pi_1}- \ket{\pi_2})/\sqrt{2}$ we require a further phase-flip described below.

For definiteness we choose the propagation direction of the microwave pulse perpendicular to the chain.
Within the dipole approximation the interaction of atom $n$ with the microwave pulse is given by
\begin{equation}
W_n(t)=- \vek{\mu}_n \vek{\mathcal{E}}(t) 
\end{equation}
with the dipole operator $\vek{\mu}_n$ and electric field $\vek{\mathcal{E}}(t)$.

Since our target initial electronic state is essentially located on two atoms, in the following we discuss the microwave excitation exemplarily for a dimer. 
The extension to larger systems can be easily done. In this section, we enlarge our essential state space beyond $\ket{\pi_1}$ and $\ket{\pi_2}$ to include also  the ``ground state'' $\ket{ss}$ and the doubly excited state $\ket{\{p,m_1\}\{p,m_2\}}$.
In this basis the Hamiltonian of the system can be written as
\begin{equation}
H=\left[
\begin{array}{cccc}
2 E_s & \bv{W}_1(t) &   \bv{W}_2(t) &0 \\
 \bv{W}_1^{\dagger}(t) &( E_s+E_p){\openone}_3& V_{ps}& \underline{W}_2(t)\\
\bv{W}_2^{\dagger}(t)& V_{ps}& (E_s+E_p){\openone}_3 &\underline{W}_1(t)\\
0& \underline{W}_2^{\dagger}(t) &   \underline{W}_1^{\dagger}(t) & 2 E_p{\openone}_9
\end{array}
\right]
\end{equation}
with $V_{ps}$ given by \eref{Vps} and $E_s$ and $E_p$ denoting the energies of the respective Rydberg states. Further, ${\openone}_n$ denotes a $n\times n$ unit-matrix, 
$\bv{W}_j(t)$ a $3\times 1$ vector and $\underline{W}_j(t)$ a $3\times 9$ matrix. The components $m$ of $\bv{W}_j(t)$ are given by $(\tilde{W}_j(t))_m=\bra{ss}W_j(t)\ket{\{p,m\}s}$, and similarly $\underline{W}_j(t)$ has elements given by $\bra{s \{p,m_2\}}W_1(t)\ket{\{p,m_1\}\{p,m_2\}}$

We now take the microwave to be linearly polarized and choose our quantization axis $\bv{\hat{z}}$ in the direction of the polarization, i.e.\ ${\vek{\mathcal{E}}}(t)=\mathcal{E}(t) \bv{\hat{z}}$.
Similar as in \Sref{Vdipdip} we get for the matrix elements
\begin{equation}
\bra{ss}\vek{\mu}_1 \hat{z} \ket{\{p,m\}s}=\frac{d_{\nu 0, \nu 1}}{\sqrt{3}} \delta_{m0}
\end{equation}
From the results of \sref{Vdipdip}, we then notice that microwave polarisation along the chain (see \eref{Vps1}) or perpendicular to the chain (see \eref{Vps2}) leads to a de-coupling of the angular momentum state $m=0$ from the other $m$ states. Thus we can consider the reduced Hamiltonian
\begin{equation}
H=\left[
\begin{array}{cccc}
2 E_s & \Omega_1(t) &   \Omega_2(t) &0 \\
 \Omega_1(t) &( E_s+E_p)& V_{ps}& \tilde{\Omega}_2(t)\\
\Omega_2(t)& V_{ps}& (E_s+E_p) &\tilde{\Omega}_1(t)\\
0& \tilde{\Omega}_2(t) &   \tilde{\Omega}_1(t) & 2 E_p
\end{array}
\right]
\end{equation}  
with  $\Omega_n(t)=\frac{1}{\sqrt{3}}\mathcal{E}(t)d_{\nu 0, \nu 1}$ and $\tilde{\Omega}_n$ is of the order of $\Omega_n$.
It is instructive to diagonalize within the one-exciton space to obtain the ``eigenstates'' $\ket{\pm}=\frac{1}{\sqrt{2}}(\ket{\pi_1}\pm\ket{\pi_2})$ with energies $E_{\pm}=(E_s+E_p)\pm V_{ps}$.
Taking $E_s$ as zero of energy we obtain for the Hamiltonian in this basis 
\begin{equation}
H=\left[
\begin{array}{cccc}
0 & \Omega_{+}(t) &   \Omega_{-}(t) &0 \\
 \Omega_{+}(t) &E_p+V_{ps}& 0& \tilde{\Omega}_{+}(t)\\
\Omega_{-}(t)&0& E_p- V_{ps} &\tilde{\Omega}_{-}(t)\\
0& \tilde{\Omega}_{+}(t) &   \tilde{\Omega}_{-}(t) & 2 E_p
\end{array}
\right]
\end{equation}   
with $\Omega_{\pm}=\frac{1}{\sqrt{2}}(\Omega_1\pm\Omega_2)$.
Since we are dealing with identical atoms $\Omega_{-}=0$, hence the microwave couples only to the symmetric state  $\ket{+}=\frac{1}{\sqrt{2}}(\ket{\pi_1}+\ket{\pi_2})$.
Thus in order to be resonant with the transition $\ket{ss}\rightarrow \ket{+}$ we will detune the microwave by $V_{ps}$ w.r.t.\ the atomic transition frequency. 
This also means that the microwave is detuned by $2 V_{ps}$ w.r.t\ the transition from the state $\ket{+}$ to the doubly excited state $\ket{pp}$, so that the population of the doubly excited state will be strongly suppressed.

Ideally the microwave pulse should transfer all the population from the $\ket{ss}$ state to the $\ket{+}$ state and be so short that the atoms do not move appreciably during the duration of the pulse.
We have done full numerical simulations of this excitation scheme for three Lithium atoms, and we found that pulses of few nanoseconds duration can be used to  achieve this goal. 

It also is of interest to access the aggregate eigenstate $\ket{-}$. We will actually focus on dynamics arising from a $\ket{-}$-type initialstate and interactions with $\eta=1$ throughout this article, since in that case the smaller energetic separation between the totally repulsive adiabatic state and its energetic neighbour leads to more interesting non-adiabatic effects. This scenario was also considered in our previous work \cite{cenap:motion,wuester:cradle}.

Since $\ket{-}$ does not directly couple to the linear polarised microwave, as argued above, this requires a second state preparation step in which e.g.~the phase of the $\ket{\pi_2}$ component of the quantum state is inverted. This can be achieved using a Rabi-$2\pi$ laser pulse, which is resonant on the transition from $\ket{\nu p}$ to e.g.\ the absolute ground-state $\ket{2s}$ and spatially focussed to only interact with atom two \cite{kruse:siteselectiveaddress}. 

%%%%%%%%%%%%%%%%%%%%%%%%%%%%%%%%%%%
\subsection{Dynamical methods}
%%%%%%%%%%%%%%%%%%%%%%%%%%%%%%%%%%%
\label{methods}

Up to this point we have introduced the Rydberg aggregate as an ensemble of alkali atoms with parameters chosen to enable a description of collective excitations in terms of Frenkel excitons, and explained how the atoms can be brought into the required internal electronic states. To form a \emph{flexible} Rydberg aggregate, we further wish to include motion of the atoms. We now list different possibilities to describe this motion numerically.

%%%%%%%%%%%%%%%%%%%%%%%%%%%%%%%%%%%
\subsubsection{Exact solution: Schr{\"o}dinger's equation}
\label{sec:exact}

The full quantum-mechanical many-body problem posed by the Hamiltonian \bref{Ham_Tot} is conceptually straightforward, but becomes quickly intractable as the number of atoms $N$ is increased. For small $N$ it is however no problem to directly solve the Schr{\"o}dinger equation
\begin{eqnarray}
\label{SE}
i \pdiff{}{t}\ket{\Psi}&=H \ket{\Psi}.
\end{eqnarray}
Expanding the full wave function in electronic (diabatic) states according to $\ket{\Psi(\bv{R})}=\sum_{n=1}^{N} \phi_{n}(\bv{R}) \ket{\pi_n}$, we arrive at
\begin{eqnarray}
\label{fullSE}
i \pdiff{}{t}\phi_{n}(\bv{R})&= \sum_{m=1}^N\left[ -\frac{\nabla^2_{\bv{\Pos_m}} }{2 M}  \phi_{n}(\bv{R}) + \Vdd_{nm}(\Pos_{nm}) \phi_{m}(\bv{R}) \right].
\end{eqnarray}
We solve \eref{fullSE} for three Li atoms in order to validate the quantum-classical
methods presented further below, which in turn will then be faithfully used for longer chains.
In practise, the irrelevant centre-of-mass degree of freedom is removed from
\eref{fullSE} resulting in an effectively two-dimensional (2D) problem. This
is solved on a discrete spatial grid.

The above \emph{diabatic} representation of the wave function $\phi_{n}(\bv{R})$ allows a
straight forward propagation. To interpret the results and compare them with the quantum-classical methods, it can also be beneficial to move to the \emph{adiabatic} representation 
\begin{equation}
\ket{\Psi(\bv{R})}=\sum_{k=1}^{N} \tilde{\phi}_{k}(\bv{R})
\ket{\varphi_k(\bv{R})}.
\end{equation}
The two representations are related by 
\begin{equation}
\tilde{\phi}_{k}(\bv{R}) = \sum_{n} O_{kn}(\bv{R})  \phi_{n}(\bv{R}).
\label{trafo_dia_adia}
\end{equation}
with $ O_{kn}(\bv{R})= \braket{\varphi_k(\bv{R})}{\pi_n}$. For instance, the initial state \eref{psi_tot_ini} corresponds to $ \tilde{\phi}_{\rm rep}(\bv{R})= \prod_{n=1}^N \phi_{\rm G}(\Pos_n)$ and $\tilde{\phi}_{k}(\bv{R})=0$ for $k\neq \rm rep$ in this representation.

When analysing our results, we will not show the full $N$-dimensional nuclear/atomic wave function but focus on the more intuitive total atomic density, which is given by 
\begin{equation}
\label{eq:density}
n(R)= \sum_{j=1}^N\sum_{m=1}^N \int d^{N-1}\bv{R}_{\{j\}} |\phi_{m}(\bv{R})|^{2}.
\end{equation}
Here $\int d^{N-1}\bv{R}_{\{j\}}$ denotes integration over all but the $j$th nuclear coordinate.
The density $n(R)$ gives the probability to find an atom at position $R$.

We will assume that wave functions of different atoms never occupy the same space. For the calculations shown, this assumption turned out to be valid.

%%%%%%%%%%%%%%%%%%%%%%%%%%%%%%%%%%%
\subsubsection{Quantum-classical propagation}

When the number of atoms $N$ exceeds values where the direct quantum solution of the time-dependent Schr\"odinger equation \eref{fullSE} is tractable, we resort to mixed quantum-classical methods, namely the Ehrenfest method  \cite{Topolar:bestmethod,tully:hopping2} and Tully's fewest switching algorithm~\cite{tully:hopping2,tully:hopping}. 
In both approaches, the nuclear coordinates $\bv R$ are treated classically and an ensemble of trajectories $\bv R(t)$ is propagated in a way specified below.
In order to represent the initial nuclear wave packet, we randomise the initial positions and velocities for the trajectories according to the Wigner distribution of the initial state \bref{psi_tot_ini}. 
Since the spatial density of each atom is assumed to be  Gaussian, this simply amounts to un-correlated Gaussian spread of both, position (with standard deviation $\sigma_{0}/\sqrt{2}$) and velocities (with standard deviation $1/(\sqrt{2}\sigma_{0}M )$). 
To obtain the total atomic density $n(R)$, the positions of the atoms are binned throughout all trajectories.

The excitonic propagation is done by expanding $\ket{\Psi(\bv{R},t)}=\sum_{k=1}^{N}\tilde{c}_{k}(t) \ket{\varphi_k(\bv{R})}$,
where the complex amplitudes $\tilde{c}_{k}$ are determined by 
\begin{eqnarray}
\label{TullysEOMs}
i \pdiff{}{t} \tilde{c}_{k} &=U_{k}(\bv{R})\tilde{c}_{k} - i \sum_{q=1}^{N} \dot{\bv{R}} \cdot \bv{d}_{kq} \tilde{c}_{q}, 
\end{eqnarray}
where $U_{k}(\bv{R})$ are the adiabatic potential energy surfaces defined in \eref{exciton_state} and 
\begin{equation}
\bv{d}_{kq}  = \langle \varphi_k(\bv{R}) | \bv{\nabla}_{\bv{R}} | \varphi_q(\bv{R}) \rangle 
\end{equation}
 are the so-called non-adiabatic coupling vectors.

The two methods differ in the classical propagation method for the nuclear coordinates. In the Ehrenfest method the nuclear dynamics is determined by Newton's equations
\begin{eqnarray}
\label{Ehrenfest_nuc}
M \ddot{\bv{R}}&=-\bv{\nabla}_{\bv{R}}\bar{U}(\bv R,t)
\end{eqnarray}
with the {\it average} potential $\bar{U}(\bv R,t)=\bra{\Psi(\bv{R},t)} \HamEl(\bv{R})\ket{\Psi(\bv{R},t)}=\sum_k |\tilde{c}_{k}(t)|^2 U_k({\bv R}) $.

In contrast in Tully's method  each trajectory moves classically on a {\it single} adiabatic surface $U_{k}(\bv{R})$, except for the possibility of instantaneous jumps among the adiabatic states. 
Between jumps the classical equation of motion is
\begin{eqnarray}
\label{TullysEOMs_nuc}
M \ddot{\bv{R}}&=-\bv{\nabla}_{\bv{R}}U_{k}(\bv{R}).
\end{eqnarray}
Details on Tully's method and our numerical implementation are given in \aref{tully_impl}.
Now, when performing the average over trajectories the spreading due to the surface hopping is combined with the spreading due to different trajectories for different initial classical nuclear positions.

%%%%%%%%%%%%%%%%%%%%%%%%%%%%%%%%%%%
%%%%%%%%%%%%%%%%%%%%%%%%%%%%%%%%%%%
\section{Entanglement transport} 
%%%%%%%%%%%%%%%%%%%%%%%%%%%%%%%%%%%
%%%%%%%%%%%%%%%%%%%%%%%%%%%%%%%%%%%
\label{cradle}

In the previous section we have explained the design of a flexible Rydberg aggregate and our various methods for dynamical propagation. Hence we are ready to consider the dynamical problem introduced in \rref{wuester:cradle} in more detail. We study the effect of resonant dipole-dipole interactions on a
regular linear chain of Rydberg atoms. Initially we impose a ``deformation'' in the distances between the atoms that gives rise to
 an associated localised exciton state, which is strongly repulsive. We demonstrate a
strong correlation between the resulting exciton dynamics and the
motion of the atoms.  A combined pulse of atomic displacements
(''deformation'') and localised electronic excitation
propagates adiabatically through the chain in a manner reminiscent of Newton's cradle. 
We show that this can also be viewed as adiabatic entanglement transport, since the initial electronic state $\ket{\varphi_{\rm rep}(\bv{R},t=0)}$ is a Bell-state\cite{nielson:chuang}. To see this, we re-write the initial state
\begin{eqnarray}
\label{Bellstate}
\ket{\varphi_{\rm rep}(\bv{R},t=0)} & \approx \frac{1}{\sqrt{2}} \bigg(\ket{\pi_1} - \ket{\pi_2} \bigg) \\
&= \frac{1}{\sqrt{2}}\bigg[\ket{ps} - \ket{sp}\bigg]\otimes\ket{s\cdots s},
\end{eqnarray}
where the state in square brackets concerns the dislocated atoms, and $\ket{s\cdots s}$ the rest of the chain.
Prior to demonstrating the combined transport of displacement, excitation and entanglement, we validate the quantum-classical methods required for larger chains.

%%%%%%%%%%%%%%%%%%%%%%%%%%%%%%%%%%%
%%%%%%%%%%%%%%%%%%%%%%%%%%%%%%%%%%%
\subsection{Comparison and validation of methods} 
%%%%%%%%%%%%%%%%%%%%%%%%%%%%%%%%%%%
%%%%%%%%%%%%%%%%%%%%%%%%%%%%%%%%%%%
\label{validation}

To confirm the applicability of quantum-classical numerical treatments to the dislocated chain of \sref{general_setup}, we consider the smallest nontrivial chain, namely $N=3$. In this case it is no problem to solve the full Schr{\"o}dinger equation numerically exactly.
 We are then in a position to compare all three propagation schemes outlined in \sref{methods}, full quantum-mechanics (QM), Tully's fewest switching (Tully) and the Ehrenfest method (EF). We consider two distinct scenarios: (i)  predominantly adiabatic dynamics for the validation of the quantum-classical methods for the subsequent \sref{transport}. (ii) strongly non-adiabatic dynamics, in order to highlight the differences in propagation algorithms.

\begin{figure}[bt]
\centerline{\psfig{file=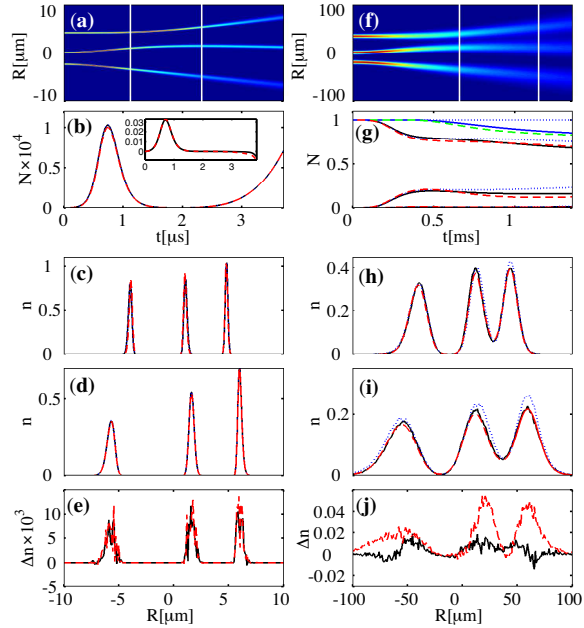, width= 0.5\columnwidth}}
\vspace*{8pt}
\caption{\label{fig:Tully_vs_QM} Comparison of QM, Tully and EF for nuclear dynamics in the
   case $N=3$ for two different parameter sets, yielding adiabatic (a-e) and
   non-adiabatic dynamics (f-j). The time evolution of the total atomic
   density $n(\bv{R},t)$ (a,f) is shown together with a comparison of Tully's
   surface hopping calculations (black solid line) with the full quantum evolution
   (red dashed line) and EF (blue dotted line) in the other panels. (b) Relative population on the energetically nearest adiabatic
 surface, $\sub{n}{mid}=\int d \bv{R} |\tilde{\phi}_{mid}(\bv{R})|^{2}$
 ($\sub{n}{mid}=|\sub{\tilde{c}}{mid}|^{2}$ in the Tully /EF algorithms), as a measure of
 the propensity of non-adiabatic transitions. The index ``mid'' is defined in the text. The inset shows the differences Tully-QM (black solid line) and EF-QM (red dashed line).
 (g) Similar to (b) but showing the total population (Tully (blue solid line), QM (green dashed line), EF (blue dotted line)) and population of all three surfaces ``rep'', ``mid'', ``att'' in descending order.
  (c,h) Spatial slice $n(x,t_{1})$, with $t_{1}$ as indicated
 by the first vertical white lines in (a,f). (d,i) Spatial slice
 $n(x,t_{2})$, with $t_{2}$ as indicated by the second vertical white lines
 in (a,f). (e,j) Difference Tully-QM and EF-QM for the density profiles at $t_{2}$ with lines as in the inset of (b).
}
\end{figure}

Scenario (i) is shown in \fref{fig:Tully_vs_QM}(a-e). We used $M=11000$~au and $\mu=1000$~au as in \rref{wuester:cradle}. The quantum mechanical probability to find an atom at a certain position predicted by QM and the corresponding semiclassical methods show perfect agreement. 
As the dynamics is almost completely adiabatic, each avoided collision between two atoms is accompanied by excitation transfer. We will highlight this in detail in the next section, where we consider longer chains.
Note that Tully and EF even perfectly reproduce the small fraction of population that has switched to the neighbouring surface
\footnote{
For the case $N=3$ there are three adiabatic surfaces, one overall repulsive (``rep''), one attractive (``att'') and one energetically between those, which we label ``mid''.

}, 
as can be seen in \frefp{fig:Tully_vs_QM}{b}. 

For scenario (ii), shown in \fref{fig:Tully_vs_QM}(f-j), we changed our parameters to $M=1800$ (hydrogen) and $\mu=200$ au. This fictitious scenario was solely chosen to increase the system's non-adiabaticity and is probably not realistic. Due to increased diffusion and collisions, we extended all three models by a phenomenological treatment of ionisation, presented and justified in \aref{ionisation}. It can be seen that in contrast to scenario (i), there are now significant transitions from the initial surface ``rep'' to ``mid''. On this surface, the trimer no longer feels an overall repulsive potential \cite{cenap:motion}. Consequently, atoms that have undergone a change of adiabatic state can approach each other closely where they ionise. This is reflected in the drop of overall population for the QM and Tully models. In contrast, atoms in the EF model always propagate according to a state averaged potential, which due to 75 \% population on the repulsive surface is still dominantly repulsive. Consequently we do not observe significant ionisation in the EF model. Despite this main difference, it can be seen that the overall state-population as well as spatial density distribution of the exact QM model is still fairly well reproduced by both Tully and EF.

The physical situation shown in \fref{fig:Tully_vs_QM}(a-e) is quite similar to that presented in the following section, except for the number of atoms.
The quality of agreement between the three disparate methods found in the case $N=3$ gives confidence that the quantum classical methods will produce reliable results also for the longer chain considered next, for which a solution of the Schr{\"o}dinger equation would no longer be feasible.

%%%%%%%%%%%%%%%%%%%%%%%%%%%%%%%%%%%
%%%%%%%%%%%%%%%%%%%%%%%%%%%%%%%%%%%
\subsection{Coupled atomic and electronic dynamics} 
%%%%%%%%%%%%%%%%%%%%%%%%%%%%%%%%%%%
%%%%%%%%%%%%%%%%%%%%%%%%%%%%%%%%%%%
\label{transport}

\begin{figure}[tb]
\psfrag{xlabel}{\small $R$  [$\mu$m]}
\centerline{\psfig{file=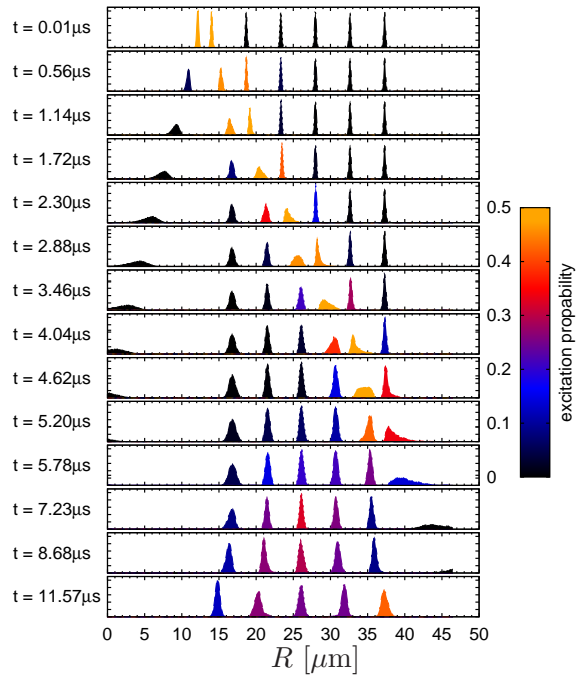,width= 0.5\columnwidth}}
\vspace*{8pt}
\caption{\label{fig:cuts} Dynamics of atomic motion and excitation transfer. Shown are snap-shots in time as labelled, of the total atomic density, as a function of spatial co-ordinate. The colour shading reflects the probability of the underlying atom to be excited, and hence demonstrates exciton transport.}
\end{figure}

\begin{figure}[bt]
\psfrag{position}{\footnotesize $R$ [$\mu$m]}
\psfrag{density}[t][1.5]{\footnotesize $n(t)$}
\psfrag{exc.prop.}[t][1.5]{\footnotesize $|c_m(t)|^2$}
\psfrag{atom}{\footnotesize $m$}
\psfrag{elecpop}[b][1.5]{\footnotesize population}
\psfrag{entangle}[b][1.5]{\footnotesize entanglement}
\psfrag{time}[t][1.5]{\footnotesize time [$\mu$s]}
\centerline{\psfig{file=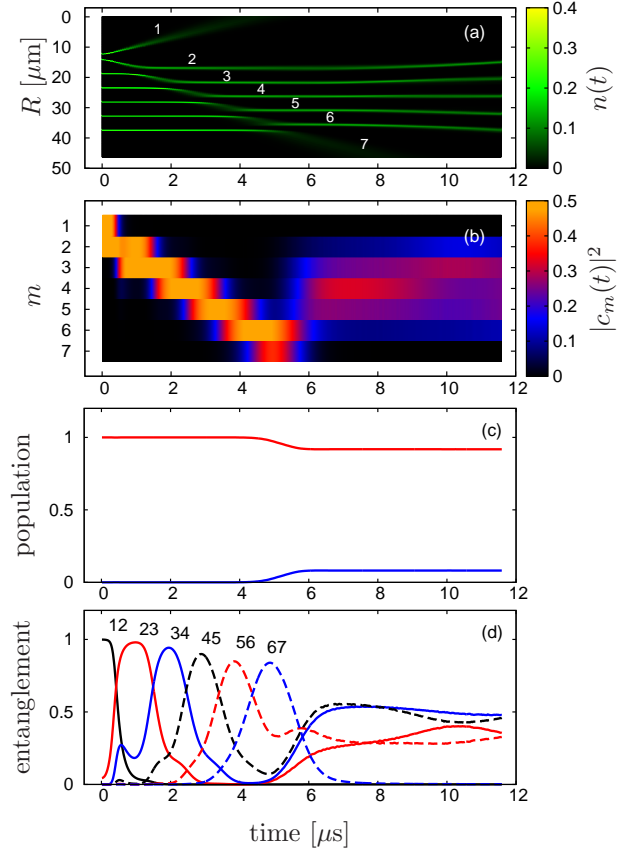,width= 0.5\columnwidth}}
\vspace*{8pt}
\caption{\label{fig:longchain} Dynamics of atomic motion and excitation transfer averaged over $10^5$ realisations. (a) Total atomic density as defined in \eref{eq:density}, obtained by binning the classical trajectories $R(t)$, see \ref{tully_impl} . (b) Diabatic populations $|c_{m}(t)|^2$ in the localised state basis $\ket{\Psi(t)} = \sum_n c_n(t) \ket{\pi_n}$. The row $m$ shows the excitation probability of atom $m$ . (c) Population on the adiabatic (eigen-) surface $rep$ (red) and the energetically next one (blue). (d) Binary entanglement (see \aref{entanglement}) $E_{n,n+1}$ for neighbouring atoms. The pair of indices $n,n+1$ is assigned to each line near its maximum. }
\end{figure}

The atomic motion and excitation transfer for a chain of $N=7$ atoms, when starting in the exciton state with highest energy (which corresponds to the fully repulsive state) is shown in \fref{fig:cuts} and \fref{fig:longchain}. As expected, initially the two close atoms strongly repel each other. When atom {\atomA} approaches atom {\atomB} these
atoms start to repel each other. Atom {\atomA} slows down and atom {\atomB}
accelerates.
In this way the momentum of atom {\atomA} is transferred through the chain to atom 7, which is reached at $t\approx 5.5 \mu$s.
Then atom 7 moves away from the remaining $N-2$ atoms, as atom 1 did already at the beginning of the evolution.
The remaining atoms form a regular chain with distance $x_0$ between the atoms and positions shifted by $x_{0}-a$ w.r.t.\ the initial position of the respective atom. 
This chain is in a repulsive state and the atoms drift very slowly apart in a manner typical for a regular chain as discussed in \cite{cenap:motion}.

Note the spreading of the initially quite localised wavepackets right from the start, for example atom 1. This is due to the initial \emph{spatial} width of the Rydberg atom distribution $\sigma_{0}$, which gets converted into strong velocity spread $\Delta v$ due to the steep slope of the dipole-dipole potential. One expects $\Delta v=(2\sigma_{0} \mu^2 / M a^4)^{1/2}$. Then estimating $\Delta x = \Delta v t$ describes the spreading of atom 1 well. Atom {\atomA} initially obtains the same large velocity spread, in the following elastic collision this is however exchanged completely against the (narrow) velocity distribution of {\atomB}. After the dislocation has traversed the chain, only the outer atoms have a considerable spread in velocity which results in a large position smear as time progresses (see \fref{fig:cuts}).
During this transfer of momentum there is negligible overlap of the spatial distributions of different atoms, even at the avoided collisions\footnote{In interpreting \fref{fig:cuts}, keep in mind that a narrow gap between the total density peaks associated with two neighbouring atoms \emph{does not} imply that the atoms approach closely: correlations between atomic positions are strong and result in the absence of actual close encounters.}.
Up to now we have restricted our discussion to the ``slow'',  ``macroscopic'' movement of the Rydberg atoms.
The interaction strength between a pair of Rydberg atoms at a distance $a=3.5 \mu$m (corresponding to the average closest distance between atoms up to the time t=$5.5\mu$s) is approximately $141$MHz, which corresponds to a ``transfer'' time of $0.02$ $\mu$s.
This is much faster than the time scale of motion of the atoms. 
The colour shading in \fref{fig:cuts} and \fref{fig:longchain}b visualise how the electronic excitation evolves in time. 
One sees that the excitation gets transferred such that it is always localised
on the two instantaneously nearest atoms, in accordance with the structure of
exciton eigenstates outlined in \cite{cenap:motion}.  After $5.5$ $\mu$s the momentum that was transferred through the chain kicks out the last atom, and a
well defined close proximity pair no longer exists. The exciton state then
assumes the shape for an equidistant chain, de-localised over the entire chain
(consisting of the remaining $N$-2 atoms), which subsequently slowly spreads out. However this state change is not completely adiabatic as can be seen in \fref{fig:longchain}c where the adiabatic population on the initial (repulsive) adiabatic surface together with the population on the neighbouring adiabatic surface is shown.
One clearly sees a change of population around $t=5.5\mu$s, which is the time when atom 7 starts to separate from the chain. The duration over which population transfer between the surfaces occurs corresponds to the time during which the excitation localised on atom 6 and 7 spreads over the remaining chain (see \fref{fig:cuts} and \fref{fig:longchain}b). 
This change of the adiabatic populations can be understood in a simple way:
As noted above, up to $t\approx 5 \mu$s the excitonic transfer time was much faster than the nuclear dynamics.
During the time in which atom 7 leaves the chain, however, the excitation has to delocalise over the whole remaining chain to stay in the fully repulsive adiabatic state. The distances involved in this redistribution of excitation are much larger than $a$ or $x_{0}$, hence the electronic time scale is slower, \emph{now of} the order of the nuclear motion of atom 7.
After the delocalised state is reached, the relevant nuclear dynamics becomes very slow -- the system behaves adiabatic again. We found that the magnitude of these non-adiabatic transitions increases with chain length $N$ if all other parameters are kept constant, reflecting a decrease in the energetic separation of the involved adiabatic states for larger $N$.

So far we have viewed the dynamics of excitation transport essentially as a wave spreading phenomenon on a chain whose constituents are free to move. It is possible to give the observed phenomenon a quite different twist, by considering the dynamical transport of entanglement that is linked to the excitation migration. In particular we focus on entanglement within the
subsystem comprised of the electronic state of atoms $n$ and $m$ only, denoted
by $E_{n,m}$. This subsystem can contain much less information than the full
many-body quantum state, hence entanglement therein is expected to be more
robust and simultaneously more accessible.  We summarize in \aref{entanglement} how we calculate the
relevant bipartite \emph{entanglement of formation} \cite{hill:wootters:qbits}. As can be seen in \frefp{fig:longchain}{d}, the initially
perfect entanglement between 1 and 2 is transported through the chain with
only minor losses up to the point where the final atom leaves the chain \footnote{We have verified that both methods outlined previously give the
	same entanglement evolution for the case $N=3$.}. 
Then the exciton state de-localizes over the entire chain, with a resulting drop of
\emph{bipartite} entanglement. A comparison of panels (b) and (d) of \fref{fig:longchain} makes it apparent that entanglement is here a direct consequence of coherent, delocalised excitation: Whenever the diabatic population on \emph{both} members of a neigbouring atom pair is large, so is the mutual entanglement. 

%%%%%%%%%%%%%%%%%%%%%%%%%%%%%%%%%%%
%%%%%%%%%%%%%%%%%%%%%%%%%%%%%%%%%%%
\section{Parameter dependence of the entanglement transport} 
%%%%%%%%%%%%%%%%%%%%%%%%%%%%%%%%%%%
%%%%%%%%%%%%%%%%%%%%%%%%%%%%%%%%%%%
\label{survey}
In the following we will investigate how the coupled excitonic and nuclear dynamics depends on the mass $M$ of the atom, the magnitude $\mu$ of the used transition  dipole and on the absolute initial positions of the atoms.
These dependencies can be in principle be studied in experiment. 
In addition we will also investigate changes of the functional form of the long-range interaction, which is more of general theoretical interest. 
 As in the previous section, for the following calculations we will use and compare the two mixed quantum classical methods.

\begin{figure}[tbp]
\psfrag{t/t0}{\footnotesize $t/t_0$}
\psfrag{E12}{\footnotesize $E_{12}$}
\psfrag{E23}{\footnotesize $E_{23}$}
\psfrag{E34}{\footnotesize $E_{34}$}
\psfrag{E45}{\footnotesize $E_{45}$}
\psfrag{E56}{\footnotesize $E_{56}$}
\psfrag{mu1}{\footnotesize $\mu=1000$ au}
\psfrag{mu2}{\footnotesize $\mu=1500$ au}
\psfrag{mu4}{\footnotesize $\mu=2000$ au}
\psfrag{mu10}{\footnotesize $\mu=10000$ au}
\center
\includegraphics[width=0.4\columnwidth]{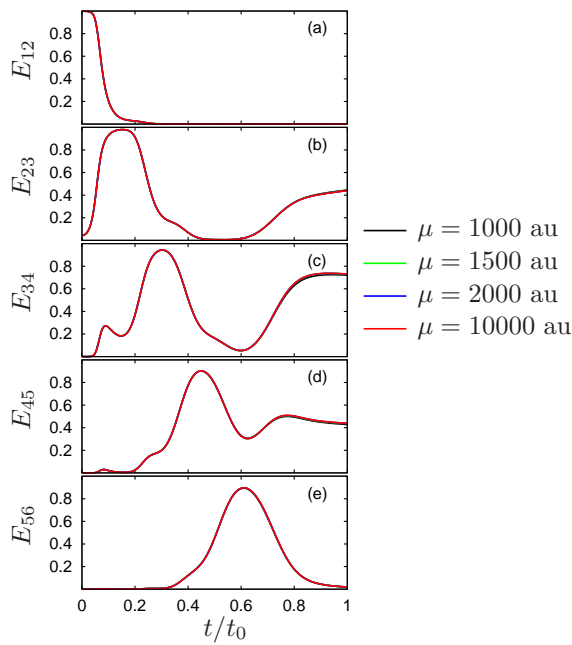}
\caption{\label{fig_mu_scal} Entanglement $E_{nm}(t)$ between neighbouring atoms $n$ and $m$ for different values of the transition dipole moment $\mu$  calculated with Tully and Ehrenfest.  For each $\mu$ the time is scaled with $t_0=t_0(\mu)=T/\mu$ with $T=6.44 \mu$s. The curves for different $\mu$ are indistiguishable. }
\end{figure}

Since we focus on dynamics which is more or less adiabatic, the motion of the atoms is approximately governed by
\begin{equation}
\label{eq_motion_rep}
M \frac{\partial^2}{\partial t^2}R=-\nabla_{R} U_{\rm rep}(R)=
-\nabla_{R}\frac{\mu^2}{R^3}=3 \frac{\mu^2}{R^4},
\end{equation}
which is equivalent to $\frac{\partial^2}{\partial t^2}R=3 \frac{\mu^2}{M R^4}$. If we now scale
\begin{eqnarray}
\label{R_scal}
R_{\lambda}&=\lambda R,\\
\label{M_scal}
M_{\beta}&=\beta M,\\
\label{mu_scal}
\mu_{\gamma}&=\gamma \mu,
\end{eqnarray}
we see that \eref{eq_motion_rep} remains invariant, if time is also scaled by
\begin{equation}
\label{tau_scale}
\tau= t \gamma^{-1} \beta^{1/2} \lambda^{3/2}.
\end{equation}
This means, for example, that for doubled transition dipole moment $\mu$ one expects the dynamics to be twice as fast, but otherwise unchanged.
This is confirmed in \fref{fig_mu_scal} where entanglement transport for various transition dipole strengths is shown, each scaled by \eref{tau_scale}.

\begin{figure}[tbp]
\center
\psfrag{t/t0}{\footnotesize $t/t_0$}
\psfrag{E12}{\footnotesize $E_{12}$}
\psfrag{E23}{\footnotesize $E_{23}$}
\psfrag{E34}{\footnotesize $E_{34}$}
\psfrag{E45}{\footnotesize $E_{45}$}
\psfrag{E56}{\footnotesize $E_{56}$}
\psfrag{dist2}{\footnotesize $a=2\ \mu$m }
\psfrag{dist2.5}{\footnotesize $a=2.5\ \mu$m }
\psfrag{dist3}{\footnotesize $a=3\ \mu$m }
\psfrag{dist4}{\footnotesize $a=4\ \mu$m }
\psfrag{dist4}{\footnotesize $a=5\ \mu$m }
\psfrag{dist8}{\footnotesize $a=8\ \mu$m }
\includegraphics[width=0.5\columnwidth]{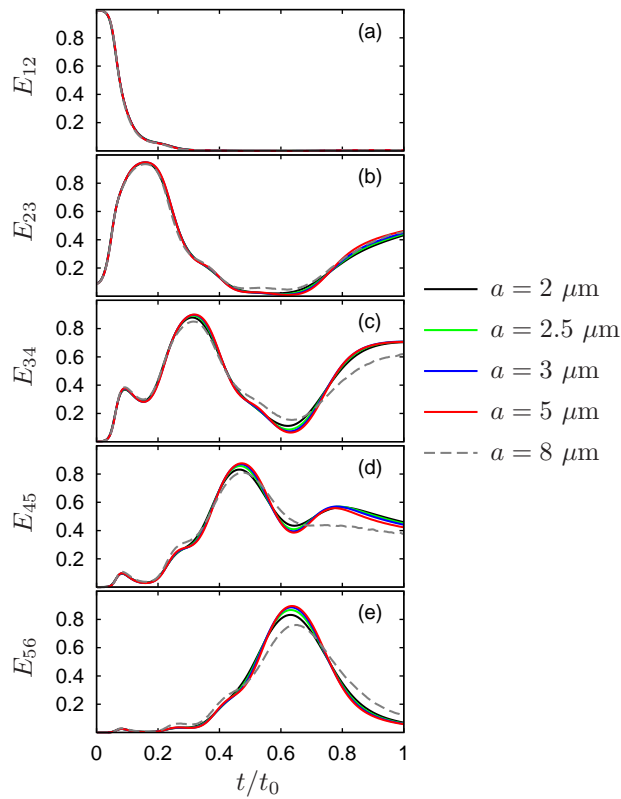}
\caption{\label{fig_distances_scale}As \fref{fig_mu_scal} but now for scaled distances $\lambda \times 2 \mu$m and constant $a/x_0$. Here the width $\sigma_0$ of the initial nuclear wave-function is kept constant. The time is given in units of  $t_0 = t_0(\lambda) = T\lambda^{5/2}$ with T = $6.44 \mu$s.}
\end{figure}

Consider next the distance dependence.
Numerical calculations are shown in \fref{fig_distances_scale}.
Here we have kept the ratio $a/x_0$ constant and scaled the distances between the atoms according to \eref{R_scal}, however we did not scale the the width $\sigma_{0}$ of the initial nuclear wavefunction. 
One sees that the overall dynamics obeys the scaling  \eref{tau_scale}, however there are slight differences in the magnitude of the entanglement.
These are due to the different relative width of the initial Gaussian.
If we scale also $\sigma_{0}$ we obtain perfect agreement as in \fref{fig_mu_scal}.

\begin{figure}[bt]
\psfrag{t/t0}{\footnotesize $t/t_0$}
\psfrag{E12}{\footnotesize $E_{12}$}
\psfrag{E23}{\footnotesize $E_{23}$}
\psfrag{E34}{\footnotesize $E_{34}$}
\psfrag{E45}{\footnotesize $E_{45}$}
\psfrag{E56}{\footnotesize $E_{56}$}
\psfrag{m0.06tully}[t][1.7]{\footnotesize $\beta=0.06$}
\psfrag{m0.1tully}[t][1.7]{\footnotesize $\beta=0.10$}
\psfrag{m1tully}[t][1.8]{\footnotesize $\ \ \ \beta=1.00$}
\psfrag{m1.49tully}[t][1.7]{\footnotesize $\beta=1.49$}
\psfrag{m0.06ehren}{\footnotesize $ $}
\psfrag{m0.1ehren}{\footnotesize $ $}
\psfrag{m1ehren}{\footnotesize $ $}
\psfrag{m1.49ehren}{\footnotesize $ $}
\centerline{\psfig{file=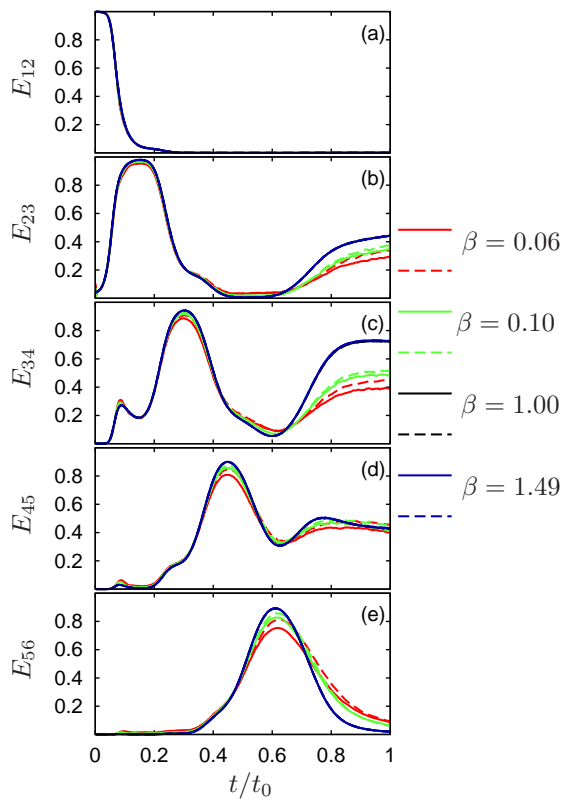, width= 0.5\columnwidth}}
\caption{\label{vgl_mass_ent_tul_ehr.eps}As \fref{fig_mu_scal} but now for different masses $M=\beta M_{\rm Li}$.  Tully's surface hopping algorithm (solid lines), Ehrenfest method (dashed lines).  The time is given in units of the mass-dependent time $t_0 = t_0(\beta) = T\sqrt{\beta}$ with T = $6.44 \mu$s. For $m=1$ and $m=1.49$ all lines  are indistiguishable.}
\end{figure}

The dependence on the mass is demonstrated in \fref{vgl_mass_ent_tul_ehr.eps}.
Again one sees that the scaling relation \eref{tau_scale} is basically fulfilled, however, 
there are quite large changes in the absolute value of entanglement when masses or calculation methods are varied, especially after the delocalisation of the excitation (see \sref{transport}). 
These differences appear pronounced for very small masses\footnote{The green curve corresponds roughly to hydrogen.}.

%%%%%%%%%%%%%%%%%%%%%%%%%%%%%%%%%%%%%%%%%%%%%%555

\begin{figure}[bt]
\psfrag{massrelLi}{\footnotesize $\beta$}
\centerline{\psfig{file=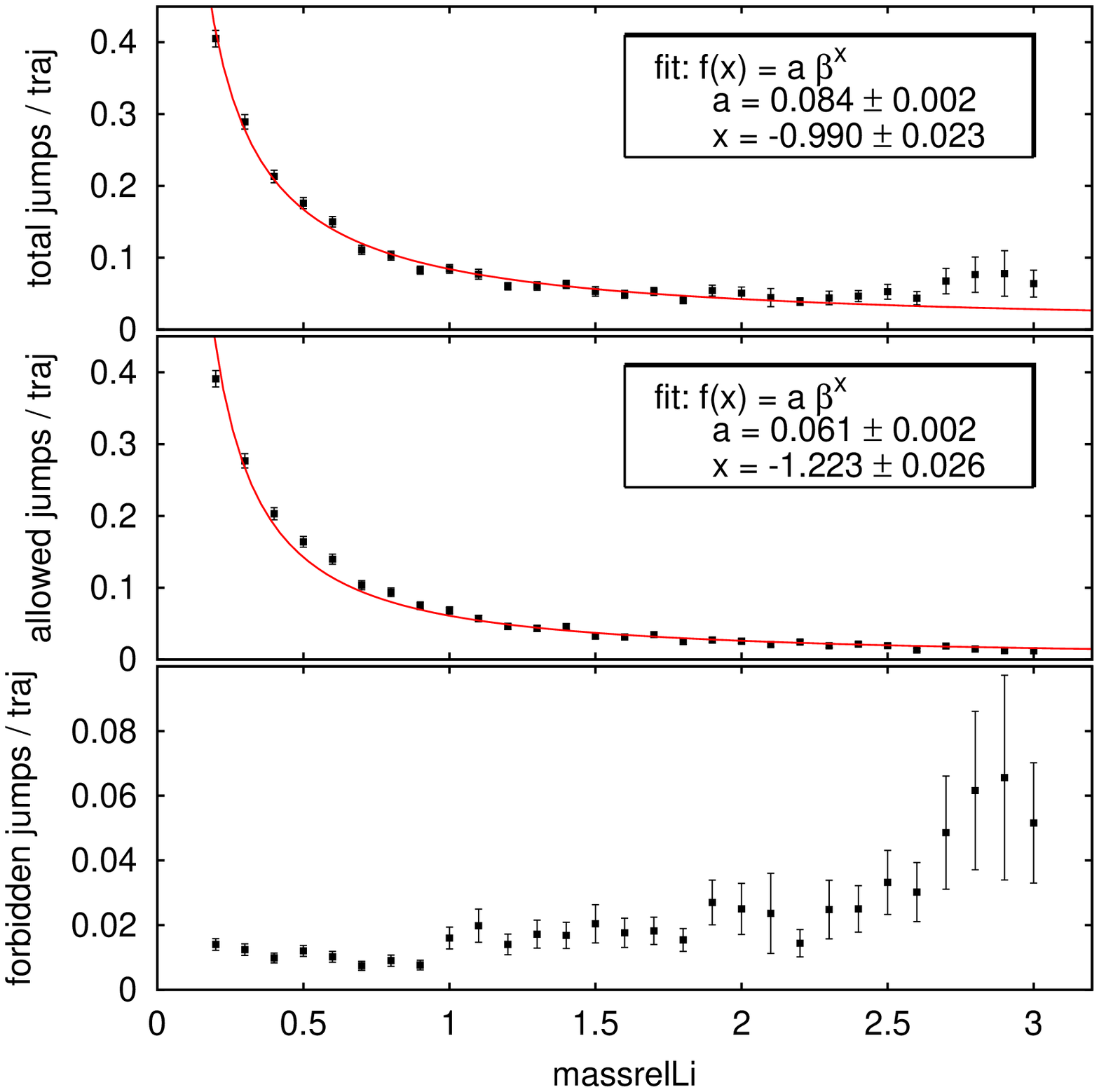, width= 0.5\columnwidth}}
\caption{\label{jumps_mass_scaling.eps} Number of jumps invoked in Tully's algorithm as function of the mass. Here $\beta = M/M_{Li}$ is the mass expressed in units of the mass of Lithium. Black squares are the numerical data (with statistical errors) and the red lines are power law fits.}
\end{figure}

To understand the deviations, the number of jumps invoked in Tully's method is shown as a function of mass in \fref{jumps_mass_scaling.eps}. As expected for smaller masses  the dynamics becomes more non-adiabatic and the number of jumps strongly increases\footnote{A similar plot but with the number of jumps reduced by roughly a factor of ten is obtained when starting from the symmetric initial electronic state, due to a larger energy-separation between neighbouring surfaces.}.

%%%%%%%%%%%%%%%%%%%%%5
%%%%%%%%%%%%%%%%%%%%%%%%%%%%%%%%%

From the above analysis one sees that due to an increasing number of jumps there are differences in the dynamics for different masses.
 However, in practice these differences are quite small as is demonstrated in \fref{vgl_mass_ent_tul_ehr.eps}.
The overall shape of the curves stays the same, only the timescales for the total dynamics changes.
Nevertheless one clearly sees that with smaller mass, where the dynamics becomes more non-adiabatic, there is also a larger loss of entanglement.

It is instructive also to compare Tully's surface hopping approach with the much simpler (and faster) Ehrenfest method.
For larger masses, in \fref{vgl_mass_ent_tul_ehr.eps}, the results obtained from Tully's surface hopping method and the Ehrenfest-average-potential method do not differ, since only a few  transitions between the adiabatic states occur.
For lighter masses, where more transitions occur, the methods start to deviate, since the dynamics on the averaged potential  in the Ehrenfest-method differs from that stemming from Tully's algorithm.

%%%%%%%%%%%%%%%%%%%%%%%%%%%%%%%%%%%
\subsection{Dependence on the interaction potential}
%%%%%%%%%%%%%%%%%%%%%%%%%%%%%%%%%%%

The physical transition dipole-dipole interaction between the states $\ket{\pi_n}$ scales with $1/R^3$, where $R$ is the distance between two Rydberg atoms. 
In the following we will investigate (hypothetical) resonant energy transfer interactions with power law dependence on the distance
\begin{equation}
V_{nm} = -\mu^2/R_{nm}^\alpha.
\end{equation}
As already shown in \fref{exciton_localisation}, to obtain the initial state
$\ket{\psi_{ini}}\approx(\ket{\pi_1} - \ket{\pi_2})/\sqrt{2}$, the ratio  $a/x_0$ has to be decreased for decreasing $\alpha$.
To fix the $\alpha$ dependence of $a/x_0$ by keeping the ratio of the interaction energy between atom 1 and 2 to that between atom 2 and 3 the constant compared to the case $\alpha=3$.
Parameters selected in this manner are shown in \fref{exciton_localisation} as solid black bars. 

To have comparable dynamics for different exponents $\alpha$, we further fix the values of $a$ and $x_0$. 
 We choose the $\alpha$ dependent initial distance  $a_{\alpha}$ by requiring $V_{12}(a_\alpha) \equiv V_{12}(a_3)$, which leads to $a_\alpha = (a_3)^{3/\alpha}$.  
Furthermore we have to adapt the width of the initial nuclear wave packet to the new distance. 
Our variance $\sigma_{\alpha}$ of the Gaussian position distribution is determined by requiring that the corresponding spread of initial potential energies is roughly independent of $\alpha$. 
This leads to the condition $\sigma_\alpha \approx \frac{3}{\alpha}\; a_3^{\frac{3}{\alpha}-1}\; \sigma_3$ .

\begin{figure}[bt]
\psfrag{t/t0}{\footnotesize $t/t_0$}
\psfrag{E12}{\footnotesize $E_{12}$}
\psfrag{E23}{\footnotesize $E_{23}$}
\psfrag{E34}{\footnotesize $E_{34}$}
\psfrag{E45}{\footnotesize $E_{45}$}
\psfrag{E56}{\footnotesize $E_{56}$}
\psfrag{a2.5}{\footnotesize $\alpha=2.5$}
\psfrag{a2.8}{\footnotesize $\alpha=2.8$}
\psfrag{a3}{\footnotesize $\alpha=3$}
\psfrag{a4}{\footnotesize $\alpha=4$ }
\psfrag{a6}{\footnotesize $\alpha=6$ }
\centerline{\psfig{file=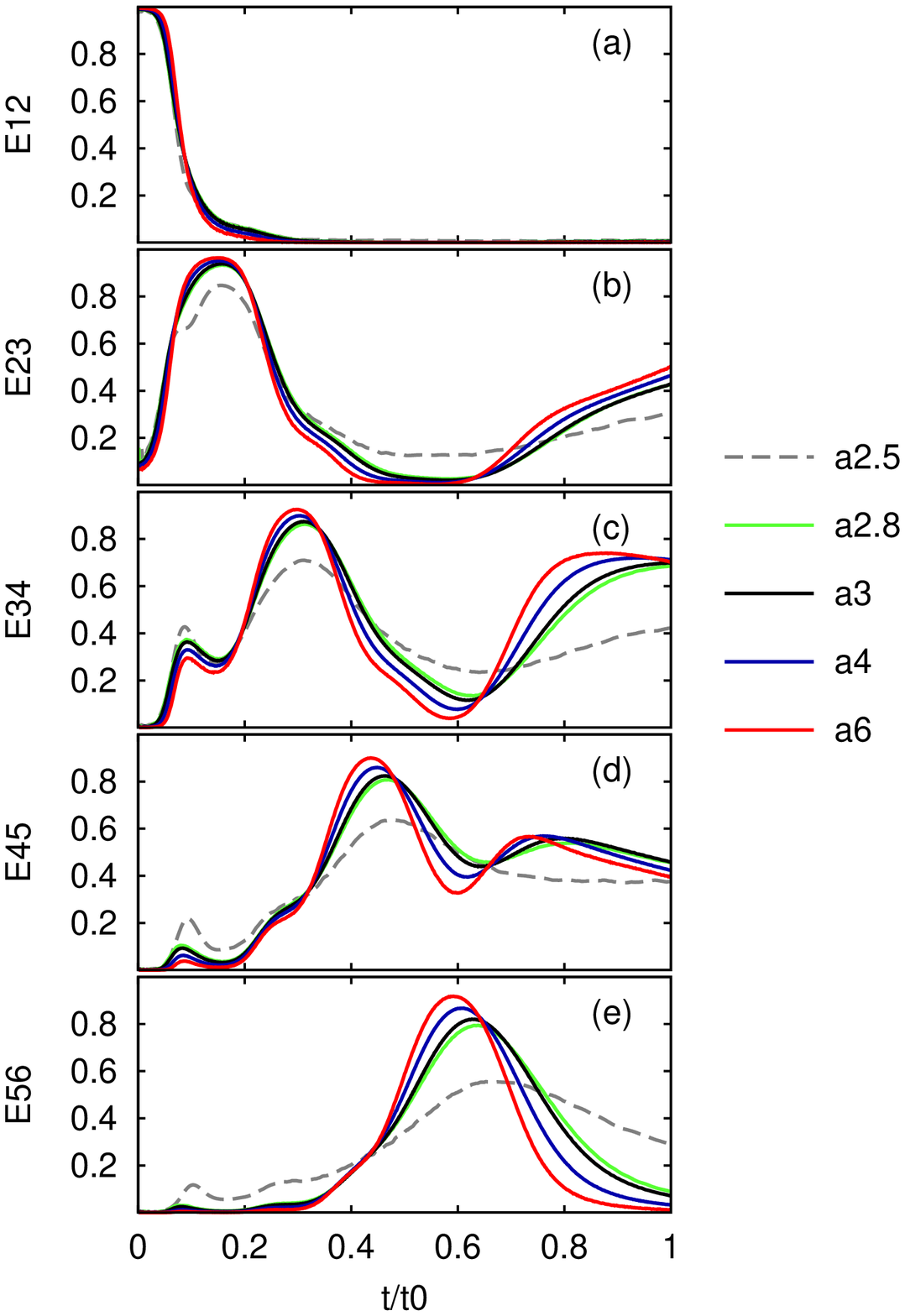, width= 0.45\columnwidth}}
\caption{\label{vgl_alpha_ent_tul_ehr}As \fref{fig_mu_scal} but now for different exponents $\alpha$. The time is scaled according to \eref{t_scal_alpha}.}
\end{figure}
%%%%%%%%%%%%%%%%%%%%%%%%

In \fref{vgl_alpha_ent_tul_ehr} the entanglement between neighbouring atoms is shown for different $\alpha$ and $M=M_{\rm Li}$.
The time is in units of $t_0(\alpha)$, which is different for each $\alpha$ 
\begin{eqnarray}
\label{t_scal_alpha}
t_0(\alpha) = \frac{\alpha}{3}\;\sqrt{(a_3 d_3)^{1-\frac{3}{\alpha}}}\; T
\end{eqnarray}
with with T = $6.44 \mu$s. One sees that in these scaled time units, for $\alpha\ge 3$, the transport of entanglement is more or less independent of $\alpha$. 
For smaller $\alpha$, however, the entanglement transport is strongly reduced.

%%%%%%%%%%%%%%%%%%%%%%%%%%%%%%%%%%%
%%%%%%%%%%%%%%%%%%%%%%%%%%%%%%%%%%%
\section{Conclusions and Outlook} 
%%%%%%%%%%%%%%%%%%%%%%%%%%%%%%%%%%%
%%%%%%%%%%%%%%%%%%%%%%%%%%%%%%%%%%%
\label{conclusion}

We have demonstrated a strong connection between the motion of a chain (aggregate) of Rydberg atoms and the coherent propagation of a single electronic excitation within the chain. Adiabatic transport ensures that the excitation remains spatially localised near a dislocation passing through the chain. Our results were obtained with Tully's surface hopping method~\cite{tully:hopping2,tully:hopping} and the Ehrenfest method~\cite{Topolar:bestmethod,tully:hopping2}, both of which we vindicate by comparison with exact quantum calculations for a smaller model system with similar dynamics. A key feature of our setup is that the initial state is a repulsive electronic eigenstate of the chain.

If the system was not prepared in an electronic eigenstate but in a state where the excitation is localised on a single atom, one would find a fast excitation transfer similar to that described e.g.~in \cite{RoEiWo09_058301_,Ei11_33_}. If free motion is added to such a scenario, however, those parts of the population that necessarily initially reside on an an attractive surface can lead to fast collisions of light atoms. In that case the dynamics could be treated with the Tully algorithm, however the Ehrenfest method would fail (e.g.\ for the dimer there would be no movement at all). 

We wish to contrast the results of the present study with those obtained in \cite{amthor:vanderwaals} for van-der-Waals interactions.
For the case of repulsive van-der-Waals interaction one finds similar trajectories for the movement of the atoms, showing for example Newton's cradle like transfer of a dislocation through an atomic chain. The crucial difference to the dipole-dipole coupling presented here is the excitation energy transfer involved in the latter. Even more important is that the dynamics in the dipole-dipole case depends on the electronic state. An even stronger contrast to van-der-Waals is found for initial states other than the fully repulsive one treated here. Even mixed, partially attractive, partially repulsive dynamics is possible as shown by Ates {\it et al.} \cite{cenap:motion}. Such dynamics would arise from scenarios presented here only if a sufficient fraction of the atoms has undergone non-adiabatic transitions to other potential surfaces (as in our \fref{fig:Tully_vs_QM}(d)).

In this article we have exclusively studied free Rydberg atoms. When the atoms are trapped, the dipole-dipole forces will induce oscillations of atoms in the traps, which in turn lead to oscillating couplings and again to a correlation between the motion of the exciton and the motion of the Rydberg atoms. This gives rise to the well known Davydov-soliton \cite{davydov:soliton,weidlich:pulse}. Constructing large ``crystals'' of $N$ Rydberg atoms is problematic since each atom has a finite life-time $\tau$, leading to an even shorter life-time of the crystal of $\tau/N$. For the parameters used in our simulations of \sref{cradle}, we expect the lifetime to be sufficient. Therefore it would be advantageous to map the strong dipole-dipole interaction in the Rydberg state to the ground state using off-resonant laser dressing techniques \cite{wuester:dressing}. Using this technique and a ring-geometry, it is even possible to use (dressed) Rydberg aggregates for the study of near conical-intersection dynamics \cite{wuester:CI}.

Experimentally, observables as shown in \fref{fig:longchain} could be monitored using techniques for the simultaneous position and state 
measurement of Rydberg atoms \cite{ditzhuijzen:posstate}. In our particular system the presence of entanglement can then be directly inferred 
from the state populations.

%%%%%%%%%%%%%%%%%%%%%%%%%%%%%%%%%%%
%%%%%%%%%%%%%%%%%%%%%%%%%%%%%%%%%%%
\appendix
%%%%%%%%%%%%%%%%%%%%%%%%%%%%%%%%%%%
%%%%%%%%%%%%%%%%%%%%%%%%%%%%%%%%%%%
%%%%%%%%%%%%%%%%%%%%%%%%%%%%%%%%%%%
%%%%%%%%%%%%%%%%%%%%%%%%%%%%%%%%%%%
\section{Entanglement of formation}
%%%%%%%%%%%%%%%%%%%%%%%%%%%%%%%%%%%
%%%%%%%%%%%%%%%%%%%%%%%%%%%%%%%%%%%
\label{entanglement}

The ``entanglement of formation''
\cite{hill:wootters:qbits,wootters:mixed} is an entanglement measure for
bipartite quantum states, also applicable to mixed states. For a pure
state it equals $1$ for perfect entanglement and it is $0$ for a separable state.

We calculate this entanglement measure in the following way: First consider the reduced density matrix describing the electronic state when the atomic positions are traced out
\begin{eqnarray}
\hat{\sigma}&=\sum_{n,m}\sigma_{nm} \ket{\pi_{n}}\bra{\pi_{m}},
\label{red_el_Dm}
\end{eqnarray}
with
\begin{eqnarray}
\sigma_{nm}=\left\{ \begin{array}{c}
\int d^N \bv{R}\:\:  \phi^*_{n}(\bv{R}) \phi_{m}(\bv{R}) \hspace{1cm}
\hfill {\rm QM}\\
\\
\overline{c_{n}^* c_{m}}\hfill {\rm Tully/Ehrenfest} .
\end{array}
\right.
\label{red_el_Dm_ME}
\end{eqnarray}
The first expression holds for the full quantum calculations, the second
for the quantum-classical methods. In the latter case $\overline{\cdots}$
denotes the trajectory average and $c_{n}=\sum_{k} O_{nk}^T \tilde{c}_{k}$ are the coefficients in the diabatic basis with  $O_{nk}$ defined in  \eref{trafo_dia_adia}.

From \eref{red_el_Dm_ME} we construct the binary reduced electronic density
matrix of atoms $a$ and $b$
\begin{equation}
\hat{\beta}_{ab}={\mbox{Tr}}^{\{a,b\}}\big[\hat{\sigma}\big].
\label{beta}
\end{equation}
The symbol ${\mbox{Tr}}^{\{a,b\}}\big[\cdots\big]$ denotes the trace over the
electronic states for all atoms other than $a$, $b$.
Recall that in the present approach each atom is described by a two
 level system as discussed in \sref{general_setup} to \ref{initialstate}.
With our labels for those two states, $\ket{s}$ and $\ket{p}$, the trace appearing in \eref{beta} is over the Hilbert space spanned by
the basis $\{\ket{n_1}\cdots\ket{n_N}, \ \ n_j \in s,p   \}$

The remaining reduced subspace of atoms $a$ and $b$  is spanned by $\ket{pp}$,
$\ket{ps}$,  $\ket{sp}$,  $\ket{ss}$. Since the states $\ket{\pi_n}$ appearing in \eref{red_el_Dm} only
 contain a single excitation $p$, all matrix elements
 of $\hat{\beta}_{ab}$ involving $\ket{pp}$ vanish.
In the reduced basis one finally has
\begin{eqnarray}
\hat{\beta}_{ab}=\left(
\begin{array}{cccc}
0 & 0 & 0 & 0 \\
0 & \sigma_{aa} &  \sigma_{ab} & 0 \\
0 &  \sigma_{ab}^* &  \sigma_{bb} & 0 \\
0 & 0 & 0 & \sum_{c\neq \{a,b\}} \sigma_{cc}  \\
\end{array}
\right).
\label{betafull}
\end{eqnarray}
From this we construct
\begin{eqnarray}
S_{ab}&=\sqrt{\sqrt{\hat{\beta}_{ab}} \hat{\beta}_{ab}^*
\sqrt{\hat{\beta}_{ab}}},
\\
C_{ab}&=\max(0,2\lambda_{ab} - \mbox{Tr} S_{ab}),
\end{eqnarray}
With the further definitions
\begin{eqnarray}
H(x)&=-[x \log_{2} x + (1-x) \log_{2}(1-x)],
\\
{\cal E}(x)&=H(1/2 + \sqrt{1-x^2}/2),
\end{eqnarray}
where $\lambda_{ab}$ denotes the largest eigenvalue of $S_{ab}$, we can
finally obtain the binary entanglement of the electronic
states of atoms $a$ and $b$
\begin{eqnarray}
E_{ab}&={\cal E}(C_{ab}).
\end{eqnarray}
For further details we refer to \cite{hill:wootters:qbits,wootters:mixed}.

%%%%%%%%%%%%%%%%%%%%%%%%%%%%%%%%%%%
%%%%%%%%%%%%%%%%%%%%%%%%%%%%%%%%%%%
\section{Tully's surface hopping}
%%%%%%%%%%%%%%%%%%%%%%%%%%%%%%%%%%%
%%%%%%%%%%%%%%%%%%%%%%%%%%%%%%%%%%%
%%%%%%%%%%%%%%%%%%%%%%%%%%%%%%%%%%%

\label{tully_impl}
%%%%%%%%%%%%%%%%%%%%%%%%%%%%%%%%%%%
The quantum mechanical dynamics governed by
\eref{TullysEOMs}
and the classical equation of motion
\eref{TullysEOMs_nuc}
are solved self-consistently.
The  atoms move on a single adabatic potential surface $k$, which however may be changed via sudden jumps to another surface $q$. 
The probability for a jump from state k to state q is given by 
\begin{equation}
\label{app_jump_prob}
g_{kq} = \max(0,\frac{b_{qk}\Delta t}{a_{kk}}),
\end{equation}
where $\Delta t$ denotes the propagation time step and 
\begin{eqnarray}
 b_{qk} & =  - 2 \: \Real(a_{qk}^* \dot{\bv{R}} \cdot \bv{d}_{qk}),  \\
a_{qk} & = c_q c_k^* \;.
\end{eqnarray}
To determine if during a time step a jump takes place we compare $g_{kq}$  with a uniform random number $\xi \in [0,1]$. 
If $\xi \leq g_{k1}$, the jump is to the surface $q=1$, if $g_{k1} < \xi \leq g_{k1}+g_{k2}$,  to $q=2$ and so forth. 
When a switch takes place the velocity $\dot{\bv{R}}$ is  adjusted in order to conserve the total amount of energy. 
This will be done in the direction of the non-adiabatic coupling vector $\bv{d}_{kq}$ as follows
\begin{equation}
\label{app_adj_vel}
\dot{\bv{R}}(t) = \dot{\bv{R}}(t-\Delta t) - \frac{\gamma_{kq} \bv{d}_{kq}}{M}.
\end{equation}
Here
\begin{eqnarray}
\label{gammaTully}
\gamma_{kq} &= \frac{\beta_{kq} \pm \sqrt{\beta_{kq}^2 + 4\alpha_{kq}(U_k-U_q)}}{2\alpha_{kq}}  ,\ \beta_{kq}\lessgtr 0
%\begin{smallmatrix}< \\ \geq \end{smallmatrix} 0 \\
%  \gamma_{kq} &= \frac{\beta_{kq}}{\alpha_{kq}} , \qquad\qquad \beta_{kq}^2 + 4\alpha_{kq}(U_k-U_q) < 0
\end{eqnarray}
with
\begin{eqnarray}
\alpha_{kq} &= \frac{1}{2M} \sum_{i=1}^N | d_{kq}^{(i)} |^2\\
\beta_{kq} &= \sum_{i=1}^N \dot R_i \cdot d_{kq}^{(i)}.
\end{eqnarray}
If the energy of the final surface $q$ is larger than that of the initial surface $k$ and the velocity reduction required is greater than the component of velocity to be adjusted, then the jump is rejected and instead of \eref{gammaTully} we use
\begin{equation}
\gamma_{kq} = \frac{\beta_{kq}}{\alpha_{kq}}.
\end{equation}
which corresponds to a reflection of the velocity component along $\bv{d}_{kq}$.
Further details about Tully's algorithm can be found e.g.\ in \cite{tully:hopping2,tully:hopping}.

As described in \cite{HaTu94_4657_} the forces  $\bv{\nabla}_{\bv{R}}U_k(\bv{R})=\bv{\nabla}_{\bv{R}}\langle \varphi_k(\bv{R}) | \HamEl(\bv{R}) | \varphi_k(\bv{R}) \rangle$ and the non-adiabatic coupling vectors $\bv{d}_{ij}$ are calculated using the Hellman-Feynman theorem. One then finds 
\begin{eqnarray}\bv{\nabla}_{\bv{R}}U_k(\bv{R})=
\langle \varphi_m(\bv{R}) | (\bv{\nabla}_{\bv{R}} \HamEl(\bv{R})) | \varphi_m(\bv{R}) \rangle .
\end{eqnarray}
and
\begin{eqnarray}
\bv{d}_{kq} = \frac{\langle \varphi_i(\bv{R}) | (\bv{\nabla}_{\bv{R}} \HamEl(\bv{R})) | \varphi_j(\bv{R}) \rangle}{U_{j}(\bv{R}) - U_{i}(\bv{R})}
\end{eqnarray}

The total density $n(\bv{R},t)$, as defined in \eref{eq:density}, is obtained through a binning of the single trajectories, which means for each time step our spatial domain is discretised into bins and if a trajectory $\bv{R(t)} = (\bv{R_1(t)} ... \bv{R_N(t)})^T$ lies within such a bin, $n(\bv{R},t)$ for that bin will be increased by one. By normalising $n(\bv{R},t)$ one obtains the probability to find a atom at a given time in a certain interval of space.

%%%%%%%%%%%%%%%%%%%%%%%%%%%%%%%%%%%
\section{Phenomenological model of ionisation}
%%%%%%%%%%%%%%%%%%%%%%%%%%%%%%%%%%%
\label{ionisation}

Our essential states model as justified in \sref{essential} is only valid while atoms do not approach each other closely. Once they do, dipole-dipole shifts of all electronic states become too large to work in a small Hilbert-space of electronic states. The most prominent consequence is collisional ionisation of Rydberg atoms \cite{amthor:vanderwaals}. 
In order to avoid excursions of our numerical propagation schemes into realms where the underlying model is invalid, we incorporate a simple, phenomenological treatment of ionisation for very close atoms.

For the quantum mechanical calculations, we employ an imaginary absorbing potential of the form
\begin{eqnarray}
\label{absorberSE}
i \pdiff{}{t}\phi_{n}(\bv{R})&=\dots -i W(\bv{R})\phi_{n}(\bv{R}).
\end{eqnarray}
into \eref{fullSE}. The shape of $W(\bv{R})$ is chosen to minimise reflection, while fully removing components of the wave function that correspond to atoms closer than an enforced minimal distance.
For the data of \fref{fig:Tully_vs_QM}(f-j), where the ionisation is most important, this distance is $2\mu$m.  For the two quantum-classical trajectory methods (EF, Tully), we incorporate the effect described by \eref{absorberSE} through a stochastic ``ionisation probability'' $2W(x)\Delta t$ in each discrete time-step of duration $\Delta t$.

Note that we do \emph{not} aim to model realistic ionisation rates, however we \emph{do} employ the same model of ionisation in all three methods (QM, EF, Tully).
Further we point out that this approach practically leads to ionisation of all $N$ atoms, even if only two atoms collided. A physically correct treatment would require a density matrix formalism, going far beyond our goal, to simply to exclude numerical data from unrealistic regions of the model.

%%%%%%%%%%%%%%%%%%%%%%%%%%%%%%%%%%%
%%%%%%%%%%%%%%%%%%%%%%%%%%%%%%%%%%%

\vspace{1cm}
\section*{References}
\bibliography{Chain_Exciton_final}

\begin{thebibliography}{10}
\providecommand{\url}[1]{\texttt{#1}}
\providecommand{\urlprefix}{URL }
\expandafter\ifx\csname urlstyle\endcsname\relax
  \providecommand{\doi}[1]{doi:\discretionary{}{}{}#1}\else
  \providecommand{\doi}{doi:\discretionary{}{}{}\begingroup
  \urlstyle{rm}\Url}\fi
\providecommand{\eprint}[2][]{\url{#2}}

\bibitem{lukin:quantuminfo}
M.~D. Lukin, M.~Fleischhauer, R.~{C\^ot\'e}, L.~M. Duan, D.~Jaksch, J.~I. Cirac
  and P.~Zoller; \emph{Dipole Blockade and Quantum Information Processing in
  Mesoscopic Atomic Ensembles}; Phys. Rev. Lett. \textbf{87} 037901 (2001).

\bibitem{urban:twoatomblock}
E.~Urban, T.~A. Johnson, T.~Henage, L.~Isenhower, D.~D. Yavuz, T.~G. Walker and
  M.~Saffmann; \emph{Observation of Rydberg blockade between two atoms}; Nature
  Physics \textbf{5} 110 (2009).

\bibitem{gaetan:twoatomblock}
A.~Ga{\"e}tan, Y.~Miroshnychenko, T.~Wilk, A.~Chotia, M.~Viteau, D.~Comparat,
  P.~Pillet, A.~Browaeys and P.~Grangier; \emph{Observation of collective
  excitation of two individual atoms in the Rydberg blockade regime}; Nature
  Physics \textbf{5} 115 (2009).

\bibitem{Greene:LongRangeMols}
C.~H. Greene, A.~S. Dickinson and H.~R. Sadeghpour; \emph{Creation of Polar and
  Nonpolar Ultra-Long-Range Rydberg Molecules}; Phys. Rev. Lett. \textbf{85}
  2458 (2000).

\bibitem{liu:ultra_long_range_2009}
I.~C.~H. Liu, J.~Stanojevic and J.~M. Rost; \emph{{Ultra-Long-Range} Rydberg
  Trimers with a Repulsive {Two-Body} Interaction}; Phys. Rev. Lett.
  \textbf{102} 173001 (2009).

\bibitem{pfau:rydberg_trimers}
V.~Bendkowsky, B.~Butscher, J.~Nipper, J.~B. Balewski, J.~P. Shaffer,
  R.~L{\"o}w, T.~Pfau, W.~Li, J.~Stanojevic, T.~Pohl and J.~M. Rost;
  \emph{{Rydberg Trimers and Excited Dimers Bound by Internal Quantum
  Reflection}}; Phys. Rev. Lett. \textbf{105} 163201 (2010).

\bibitem{anderson:resonant_dipole}
W.~R. Anderson, J.~R. Veale and T.~F. Gallagher; \emph{Resonant Dipole-Dipole
  Energy Transfer in a Nearly Frozen Rydberg Gas}; Phys. Rev. Lett. \textbf{80}
  249 (1998).

\bibitem{li_gallagher:dipdipexcit}
W.~Li, P.~J. Tanner and T.~F. Gallagher; \emph{Dipole-Dipole Excitation and
  Ionization in an Ultracold Gas of Rydberg Atoms}; Phys. Rev. Lett.
  \textbf{94} 173001 (2005).

\bibitem{noordam:interactions}
F.~Robicheaux, J.~V. Hernandez, T.~Topcu and L.~D. Noordam; \emph{Simulation of
  coherent interactions between Rydberg atoms}; Phys. Rev. A \textbf{70} 042703
  (2004).

\bibitem{amthor:vanderwaals}
T.~Amthor, M.~{Reetz-Lamour}, C.~Giese and M.~Weidem{\"u}ller; \emph{Modeling
  many-particle mechanical effects of an interacting Rydberg gas}; Phys. Rev. A
  \textbf{76} 054702 (2007).

\bibitem{schempp:poptrap}
H.~Schempp, G.~G{\"u}nter, C.~S. Hofmann, C.~Giese, S.~D. Saliba, B.~D.
  DePaola, T.~Amthor, M.~Weidem{\"u}ller, S.~Sevin{\c c}li and T.~Pohl;
  \emph{{Coherent Population Trapping with Controlled Interparticle
  Interactions}}; Phys. Rev. Lett. \textbf{104} 173602 (2010).

\bibitem{lesanovsky:manybodyspin}
I.~Lesanovsky; \emph{Many-Body Spin Interactions and the Ground State of a
  Dense Rydberg Lattice Gas}; Phys. Rev. Lett. \textbf{106} 025301 (2011).

\bibitem{nils:supersolids}
N.~Henkel, R.~Nath and T.~Pohl; \emph{Three-Dimensional Roton Excitations and
  Supersolid Formation in Rydberg-Excited Bose-Einstein Condensates}; Phys.
  Rev. Lett. \textbf{104} 053004 (2010).

\bibitem{MueBlAm07_090601_}
O.~M{\"u}lken, A.~Blumen, T.~Amthor, C.~Giese, M.~Reetz-Lamour and
  M.~Weidem\"uller; \emph{{Survival probabilities in coherent exciton transfer
  with trapping}}; Phys. Rev. Lett. \textbf{99} 090601 (2007).

\bibitem{cenap:motion}
C.~Ates, A.~Eisfeld and J.~M. Rost; \emph{Motion of Rydberg atoms induced by
  resonant dipole-dipole interactions}; New J. Phys. \textbf{10} 045030 (2008).

\bibitem{Fr30_198_}
J.~Frenkel; \emph{Zur Theorie der Resonanzverbreiterung von Spektrallinien};
  Zeitschrift f{\"u}r Physik A \textbf{59} 198 (1930).

\bibitem{Fr31_17_}
J.~Frenkel; \emph{{On the Transformation of Light into Heat in Solids. {I}}};
  Phys. Rev. \textbf{37} 17 (1931).

\bibitem{FrTe38_861_}
J.~Franck and E.~Teller; \emph{{Migration and Photochemical Action of
  Excitation Energy in Crystals}}; J. Chem. Phys. \textbf{6} 861 (1938).

\bibitem{Da62__}
A.~Davydov; \emph{{Theory of Molecular Excitons}}; McGraw-Hill (1962).

\bibitem{AmVaGr00__}
H.~van Amerongen, L.~Valkunas and R.~van Grondelle; \emph{{Photosynthetic
  Excitons}}; World Scientific, Singapore (2000).

\bibitem{Ko96__}
T.~Kobayashi, editor; \emph{{J-Aggregates}}; World Scientific (1996).

\bibitem{Bi67_1484_}
A.~Bierman; \emph{Motion of Vibrationally Coupled Excition Wave Packet in an
  Infinite 1-Dimensional Crystal}; J Chem Phys \textbf{46} 1484 (1967).

\bibitem{HaRe71_253_}
H.~Haken and P.~Reineker; \emph{The coupled coherent and incoherent motion of
  excitons and its influence on the line shape of optical absorption}; Z. Phys.
  \textbf{249} 253 (1971).

\bibitem{RoEiWo09_058301_}
J.~Roden, A.~Eisfeld, W.~Wolff and W.~T. Strunz; \emph{Influence of Complex
  Exciton-Phonon Coupling on Optical Absorption and Energy Transfer of Quantum
  Aggregates}; Phys. Rev. Lett. \textbf{103} 058301 (2009).

\bibitem{ReChAs09_184102_}
P.~Rebentrost, R.~Chakraborty and A.~Aspuru-Guzik; \emph{Non-{Markovian}
  quantum jumps in excitonic energy transfer}; J. Chem. Phys. \textbf{131}
  184102 (2009).

\bibitem{BlSi78_3589_}
A.~Blumen and R.~Silbey; \emph{{Exciton line shapes and migration with
  stochastic exciton lattice coupling}}; J. Chem. Phys. \textbf{69} 3589
  (1978).

\bibitem{BaSz87_339_}
I.~Barv\'{i}k and V.~Sz\"ocs; \emph{Coherent and incoherent exciton motion in
  the framework of the continuous-time random walk}; Physics Letters A
  \textbf{125} 339 (1987).

\bibitem{Ho66_208_}
F.~Hofelich; \emph{Die Bewegung eines Exzitons entlang eines Polymers unter dem
  Einfluß der Gitterschwingungen}; Z. Phys. B: Physik Der Kondensierten
  Materie \textbf{5} 208 (1966).

\bibitem{EiBr02_61_}
A.~Eisfeld and J.~S. Briggs; \emph{{The {J}-band of organic dyes: lineshape and
  coherence length}}; Chem. Phys. \textbf{281} 61 (2002).

\bibitem{CaChDa09_105106_}
F.~Caruso, A.~W. Chin, A.~Datta, S.~F. Huelga and M.~B. Plenio; \emph{Highly
  efficient energy excitation transfer in light-harvesting complexes: The
  fundamental role of noise-assisted transport}; The Journal of Chemical
  Physics \textbf{131} 105106 (2009).

\bibitem{sarovar_fleming:entanglement}
M.~Sarovar, A.~Ishizaki, G.~R. Fleming and K.~B. Whaley; \emph{Quantum
  entanglement in photosynthetic light harvesting complexes}; Nature Physics
  \textbf{6} 462 (2010).

\bibitem{engel_fleming:coherence_nature}
G.~S. Engel, T.~R. Calhoun, E.~L. Read, {T.-K. Ahn}, T.~Man{\v c}al, {Y.-C.
  Cheng}, R.~E. Blankenship and G.~R. Fleming; \emph{Evidence for wavelike
  energy transfer through quantum coherence in photosynthetic systems}; Nature
  \textbf{446} 782 (2007).

\bibitem{lee_fleming:coherence_science}
H.~Lee, {Y.-C. Cheng} and G.~R. Fleming; \emph{Coherence dynamics in
  photosynthesis: protein protection of excitonic coherence}; Science
  \textbf{316} 1462 (2007).

\bibitem{Collini_scholes:coherence_science}
E.~Collini and G.~D. Scholes; \emph{Coherent intrachain energy migration in a
  conjugated polymer at room temperature}; Science \textbf{323} 369 (2009).

\bibitem{LiTaGa05_173001_}
W.~H. Li, P.~J. Tanner and T.~F. Gallagher; \emph{{Dipole-dipole excitation and
  ionization in an ultracold gas of {R}ydberg atoms}}; Phys. Rev. Lett.
  \textbf{94} 173001 (2005).

\bibitem{wuester:cradle}
S.~W{\"u}ster, C.~Ates, A.~Eisfeld and J.~M. Rost; \emph{Newton's cradle and
  entanglement transport in flexible Rydberg aggregates}; Phys. Rev. Lett.
  \textbf{105} 053004 (2010).

\bibitem{asadian:motion}
A.~Asadian, M.~Tiersch, G.~G. Guerreschi, J.~Cai, S.~Popescu and H.~J. Briegel;
  \emph{Motional effects on the efficiency of excitation transfer}; New J.
  Phys. \textbf{12} 075019 (2010).

\bibitem{ScWo06__}
M.~Schwoerer and H.~Wolf; \emph{Organic Molecular Solids}; {Wiley-VCH} (2006).

\bibitem{EiKnKi09_658_}
D.~M. Eisele, J.~Knoester, S.~Kirstein, J.~P. Rabe and D.~A. Vanden~Bout;
  \emph{Uniform exciton fluorescence from individual molecular nanotubes
  immobilized on solid substrates}; Nature Nanotechnology \textbf{4} 658
  (2009).

\bibitem{Ta96__}
T.~Tani; \emph{{J-Aggregates in Spectral Sensitization of Photographic
  Materials}}; in T.~Kobayashi, editor, \emph{J-Aggregates}; World Scientific
  (1996).

\bibitem{TaSuKa06_16169_}
K.~Takechi, P.~K. Sudeep and P.~V. Kamat; \emph{Harvesting infrared photons
  with tricarbocyanine dye clusters}; J Phys Chem B \textbf{110} 16169 (2006).

\bibitem{ReSmCh91_4480_}
M.~Reers, T.~W. Smith and L.~B. Chen; \emph{{J-aggregate formation of a
  carbocyanine as a quantitative fluorescent indicator of membrane-potential}};
  Biochemistry \textbf{30} 4480 (1991).

\bibitem{DeCoRo01_653_}
V.~N. Dedov, G.~C. Cox and B.~D. Roufogalis; \emph{{Visualisation of
  mitochondria in living neurons with single- and two-photon fluorescence laser
  microscopy}}; Micron \textbf{32} 653 (2001).

\bibitem{RyAmPe02_801_}
S.~N. Rylova, A.~Amalfitano, D.-A. Persaud-Sawin, W.-X. Guo, J.~Chang, P.~J.
  Jansen, A.~D. Proia and R.-M. Boustany; \emph{The CLN3 Gene is a Novel
  Molecular Target for Cancer Drug Discovery}; Cancer Research \textbf{62} 801
  (2002).

\bibitem{Dae02_81_}
S.~D\"ahne; \emph{{Nanostrukturierte {J}-Aggregate als potenzielle
  Lichtsammelsysteme f{\"u}r Photosynthesen}}; Bunsen-Magazin \textbf{4} 81
  (2002).

\bibitem{KiDa06_20363_}
S.~Kirstein and S.~D\"ahne; \emph{J-aggregates of amphiphilic cyanine dyes:
  Self-organization of artificial light harvesting complexes}; International
  Journal of Photoenergy page 20363 (2006).

\bibitem{RoEiDv11_054907_}
J.~Roden, A.~Eisfeld, M.~Dvo\v{r}\'ak, O.~B\"unermann and F.~Stienkemeier;
  \emph{Vibronic line shapes of {PTCDA} oligomers in helium nanodroplets};
  Journal of Chemical Physics \textbf{134} 054907 (2011).

\bibitem{EnCaRe07_782_}
G.~S. Engel, T.~R. Calhoun, E.~L. Read, T.~K. Ahn, T.~Man\v{c}al, Y.~C. Cheng,
  R.~E. Blankenship and G.~R. Fleming; \emph{Evidence for wavelike energy
  transfer through quantum coherence in photosynthetic systems}; Nature
  \textbf{446} 782 (2007).

\bibitem{birkl:fortagh:microlensarrays}
G.~Birkl and J.~{Fort\'agh}; \emph{Micro traps for quantum informa- tion
  processing and presicion force sensing}; {Laser \& Photon. Rev.} \textbf{1}
  12 (2007).

\bibitem{stevens:Lidefects}
G.~D. Stevens, C.~Iu, T.~Bergeman, H.~J. Metcalf, I.~Seipp, K.~T. Taylor and
  D.~Delande; \emph{Precision measurements on lithium atoms in an electric
  field compared with R-matrix and other Stark theories}; Phys. Rev. A
  \textbf{53} 1349 (1996).

\bibitem{AmReWe07_023004_}
T.~Amthor, M.~Reetz-Lamour, S.~Westermann, J.~Denskat and M.~Weidem\'uller;
  \emph{{Mechanical Effect of van der {W}aals Interactions Observed in Real
  Time in an Ultracold {R}ydberg Gas}}; Physical Review Letters \textbf{98}
  023004 (2007).

\bibitem{kruse:siteselectiveaddress}
J.~Kruse, C.~Gierl, M.~Schlosser and G.~Birkl; \emph{Reconfigurable
  site-selective manipulation of atomic quantum systems in two-dimensional
  arrays of dipole traps}; Phys. Rev. A \textbf{81} 060308(R) (2010).

\bibitem{Topolar:bestmethod}
M.~S. Topaler, T.~C. Allison, D.~W. Schwenke and D.~G. Truhlar; \emph{{What is
  the best semiclassical method for photochemical dynamics of systems with
  conical intersections?}}; J. Chem. Phys. \textbf{109} 3321  (1998).

\bibitem{tully:hopping2}
J.~C. Tully and R.~K. Preston; \emph{{Trajectory Surface Hopping Approach to
  Nonadiabatic Molecular Collisions: The Reaction of $H^+$ with $D_{2}$}}; J.
  Chem. Phys. \textbf{55} 562 (1971).

\bibitem{tully:hopping}
J.~C. Tully; \emph{Molecular dynamics with electronic transitions}; J. Chem.
  Phys. \textbf{93} 1061 (1990).

\bibitem{nielson:chuang}
M.~A. Nielson and I.~L. Chuang; \emph{{Quantum computation and quantum
  information}}; Cambridge University Press (2000).

\bibitem{hill:wootters:qbits}
S.~Hill and W.~K. Wootters; \emph{Entanglement of a Pair of Quantum Bits};
  Phys. Rev. Lett. \textbf{78} 5022 (1997).

\bibitem{Ei11_33_}
A.~Eisfeld; \emph{Phase directed excitonic transport and its limitations due to
  environmental influence}; Chemical Physics \textbf{379} 33  (2011).

\bibitem{davydov:soliton}
A.~S. Davydov and N.~I. Kislukha; \emph{Solitary Excitons in One-Dimensional
  Molecular Chains}; phys. stat. sol. (b) \textbf{59} 465 (1973).

\bibitem{weidlich:pulse}
W.~Weidlich and W.~Heudorfer; \emph{Coherent Exciton-Phonon Pulse Propagation
  in One Dimension}; Z. Physik \textbf{268} 133 (1974).

\bibitem{wuester:dressing}
S.~W\"uster, C.~Ates, A.~Eisfeld and J.~M. Rost; \emph{Excitation transport
  through Rydberg dressing}; \eprint{http://arxiv.org/abs/1011.5483} (2010).

\bibitem{wuester:CI}
S.~W{\"u}ster, A.~Eisfeld and J.~M. Rost; \emph{Conical intersections in an
  ultracold gas}; {Phys.} Rev. Lett., in press;
  \eprint{http://arxiv.org/abs/1011.5489} (2011).

\bibitem{ditzhuijzen:posstate}
C.~{van Ditzhuijzen}, A.~Koenderink, L.~Noordam and H.~{van Linden van den
  Heuvell}; \emph{Simultaneous position and state measurement of Rydberg
  atoms}; Eur. J. Phys. D \textbf{40} 13 (2006).

\bibitem{wootters:mixed}
W.~K. Wootters; \emph{Entanglement of Formation of an Arbitrary State of Two
  Qubits}; Phys. Rev. Lett. \textbf{80} 2245 (1998).

\bibitem{HaTu94_4657_}
S.~Hammes-Schiffer and J.~C. Tully; \emph{Proton-Transfer In Solution -
  Molecular-Dynamics With Quantum Transitions}; J. Chem. Phys. \textbf{101}
  4657 (1994).

\end{thebibliography}
\bibliographystyle{journal_v4}

\end{document}